\RequirePackage[displaymath]{lineno}
\documentclass[a4paper,11pt]{article}
\pdfoutput=1 

\usepackage{jcappub}
\usepackage[T1]{fontenc} % if needed

\usepackage{aas_macros}
\usepackage{tablefootnote}
\usepackage[displaymath]{lineno}

\graphicspath{{./figs/}}

\notoc

\newcommand{\gammaRay}{$\gamma$ ray}
\newcommand{\gammaRays}{$\gamma$ rays}

\newcommand{\gammaRayHyph}{$\gamma$-ray}

\newcommand{\stool}[1]{\emph{#1}}

\newcommand{\gtmktime}{\stool{gtmktime}}

\newcommand{\irf}[1]{\texttt{#1}}

\newcommand{\fov}{{\rm Fo\kern-1ptV}}

\newcommand{\Fermi}{{\textit{Fermi}}}

\newlength{\enumindent}
\newcounter{fermicnt}
% ...
\newlength{\twocolfigwidth}
\newlength{\onecolfigwidth}
\newlength{\twothirdscolfigwidth}
\newcommand{\twopanel}[4]{%
  \begin{figure}[#1]
    \centering
    \includegraphics[width=\onecolfigwidth]{#2}\hfill%
    \includegraphics[width=\onecolfigwidth]{#3} 

    #4
  \end{figure}
}

\title{Search for 100\,MeV to 10\,GeV \boldmath\gammaRayHyph\ lines in the \Fermi\/-LAT data and implications for gravitino dark matter in the \boldmath$\mu\nu$SSM}

\author[a]{Andrea~Albert,}
\author[b,c,d]{Germ\'an~A.~G\'omez-Vargas,}
\author[e,b]{Michael~Grefe,}
\author[b]{Carlos~Mu\~noz,}
\author[f]{Christoph~Weniger,}
\author[a]{Elliott~D.~Bloom,}
\author[a]{Eric~Charles,}
\author[g]{Mario~N.~Mazziotta,}
\author[d]{and Aldo~Morselli}

\affiliation[a]{W.~W.~Hansen Experimental Physics Laboratory, Kavli Institute for Particle Astrophysics and Cosmology, Department of Physics and SLAC National Accelerator Laboratory, Stanford University, Stanford, CA 94305, USA}
\affiliation[b]{Instituto de F\'isica Te\'orica UAM/CSIC and Departamento de F\'isica Te\'orica, Universidad Aut\'onoma de Madrid, Cantoblanco, E-28049 Madrid, Spain}
\affiliation[c]{Since 1st January 2014 at: Instituto de Fis\'ica, Pontificia Universidad Cat\'olica de Chile, Avenida Vicu\~na Mackenna 4860, Santiago, Chile}
\affiliation[d]{Istituto Nazionale di Fisica Nucleare, Sezione di Roma ``Tor Vergata'', I-00133 Roma, Italy}
\affiliation[e]{Institut f\"ur Theoretische Physik, Universit\"at Hamburg, Luruper Chaussee 149, D-22761 Hamburg, Germany}
\affiliation[f]{GRAPPA, University of Amsterdam, Science Park 904, 1098XH Amsterdam, Netherlands}
\affiliation[g]{Istituto Nazionale di Fisica Nucleare, Sezione di Bari, I-70126 Bari, Italy}

\emailAdd{aalbert@slac.stanford.edu}
\emailAdd{ggomezv@uc.cl}
\emailAdd{michael.grefe@desy.de}
\emailAdd{carlos.munnoz@uam.es}
\emailAdd{c.weniger@uva.nl}
\emailAdd{elliott@slac.stanford.edu}
\emailAdd{echarles@slac.stanford.edu}
\emailAdd{marionicola.mazziotta@ba.infn.it}
\emailAdd{aldo.morselli@roma2.infn.it.}

\abstract{Dark matter decay or annihilation may produce monochromatic signals in the \gammaRayHyph\ energy range. In this work we argue that there are strong theoretical motivations for studying these signals in the framework of gravitino dark matter decay and we perform a search for \gammaRayHyph\ spectral lines from 100\,MeV to 10\,GeV with \Fermi\/-LAT data. In contrast to previous line searches at higher energies, the sensitivity of the present search is dominated by systematic uncertainties across most of the energy
    range considered.  We estimate the size of systematic effects by analysing the flux from a
    number of control regions, and include the systematic uncertainties consistently in our fitting procedure. We have not observed any significant signals and present model-independent limits on \gammaRayHyph\ line emission from decaying and annihilating dark matter. We apply the former limits to the case of the gravitino, a well-known dark matter candidate in supersymmetric scenarios.  In particular, the $R$-parity violating ``$\mu$ from $\nu$'' Supersymmetric Standard Model ($\mu\nu$SSM) is an attractive scenario in which including right-handed neutrinos solves the $\mu$ problem of the Minimal Supersymmetric Standard Model while simultaneously explaining the origin of neutrino masses.  
    At the same time, the violation of
    $R$-parity renders the gravitino unstable and subject to decay into
    a photon and a neutrino. As a consequence of the limits on line emission, $\mu\nu$SSM gravitinos with masses larger than about 5\,GeV, or lifetimes smaller than about $10^{28}$\,s, are excluded at 95\% confidence level as dark matter candidates.}

\begin{document}
%\linenumbers
\maketitle
\flushbottom

% Various sections
\section{Introduction}\label{sec:intro}

The existence of non-baryonic cold dark matter (DM) in the Universe, which
today is confirmed by a large number of observations from galactic~\cite{Sofue:2000jx} to
cosmological scales~\cite{Ade:2013zuv}, is arguably the most compelling and striking
evidence for physics outside the realm of the Standard Model of particle
physics (SM)~\cite{Munoz:2003gx,Bertone:2004pz,Bergstrom:2012fi}. 
Many extensions to the SM contain 
long-lived particles that are attractive candidates for DM. 
Typically the focus is on weakly interacting massive particles (WIMPs) like 
Kaluza--Klein DM~\cite{Servant:2002aq,Cheng:2002ej}, supersymmetric
neutralinos~\cite{Munoz:2003gx,Bertone:2004pz,Bergstrom:2012fi} or right-handed 
sneutrinos~\cite{Cerdeno:2008ep,Cerdeno:2014cda}, but
non-WIMP candidates like gravitinos 
in $R$-parity violating models such as the ``$\mu$ from $\nu$'' Supersymmetric Standard
Model ($\mu\nu$SSM) may also
constitute some, if not all of the DM~\cite{Choi:2009ng}.  These DM 
candidate particles may be observed indirectly via their annihilation (in the case of WIMPs) or 
decay (as for gravitinos) into SM particles, in particular \gammaRays.  

These \gammaRays\ may be detected by the Large Area Telescope (LAT) on board the \textit{Fermi Gamma-ray Space
Telescope} (\Fermi)~\cite{REF:2009.LATPaper}, which is exploring the \gammaRayHyph\ sky
in the energy range 20\,MeV to above 300\,GeV.
We typically expect most of the DM-induced $\gamma$-ray emission to have a
broad spectrum, which can be difficult to disentangle from astrophysical
diffuse \gammaRayHyph\ backgrounds. For a recent review of indirect DM searches
with \gammaRays\ see ref.~\cite{Funk:2013gxa}.  However, spectral lines can be produced
by the two-body decay or annihilation of DM particles into final states that include 
\gammaRays.  Several searches have been performed using \Fermi\/-LAT data for
such spectral lines~\cite{Abdo:2010nc,Vertongen:2011mu,Bloom,Fermi-LAT:2013uma,Weniger:2012tx}.  Typically
these analyses have focused on searches for lines above a few GeV, since WIMPs are
expected to be heavy ($\gtrsim 10$\,GeV). We perform a search for lines below those energies, where the
statistical errors become very small and systematic uncertainties dominate.
In this paper we argue that there are strong theoretical motivations to 
search for spectral lines from lower-mass gravitino decays.  We also discuss the systematic uncertainties 
and how we incorporated them into our final results.

By far the best studied and most popular extension of the SM is the Minimal Supersymmetric Standard Model (MSSM)~\cite{martin}. For example the MSSM solves the hierarchy problem
that causes scalar masses to diverge 
in the SM. Also,
although the LHC experiments have yet to find evidence of new particles predicted in the MSSM, we expect to 
learn more when it turns back on in 2015 and reaches 14\,TeV.  Furthermore, if $R$-parity is conserved, the lightest neutralino
of the MSSM is a viable and well-known candidate for WIMP DM.
However, the MSSM has the so-called $\mu$ problem~\cite{muproblem}. This arises
from the requirement of a supersymmetric mass term for the Higgs bosons in the
superpotential, $\mu \hat H_u \hat H_d$, where $\hat H_u$ and $\hat H_d$ are
the up and down Higgs-doublet superfields, respectively.  This bilinear term is
necessary, for example, to generate Higgsino masses in order to fulfil the
current experimental bounds on chargino masses, which imply $\mu \gtrsim 100$\,GeV.
However, the
existence 
of a Grand Unified Theory (GUT) and/or a gravitational theory with typical
scales $10^{16}$ and $10^{19}$\,GeV, respectively,
would require 
an explanation of how to obtain a small supersymmetric mass, $\mu\sim 1$\,TeV, which is necessary 
in order to reproduce the correct electroweak symmetry breaking without fine tuning. 

The $\mu\nu$SSM~\cite{LopezFogliani:2005yw}
provides a solution to the $\mu$ problem through mixing terms in the superpotential between the three right-handed 
neutrino superfields, $\hat{\nu}_i^c$, and the Higgs superfields, $\lambda_i\hat \nu^c_i \hat H_d\hat H_u$. 
These produce an effective $\mu$ term\footnote{In the $\mu\nu$SSM, the usual $\mu$ term of the MSSM is absent from the 
superpotential, and only dimensionless trilinear couplings are present. This can be achieved by the presence of a 
$Z_3$ symmetry. 
Let us emphasise that this is actually what happens in superstring constructions, where the low-energy limit is determined by the massless superstring modes. Since the massive modes have huge masses, of the order of the string scale, only the trilinear couplings for the massless modes are relevant.} when the electroweak symmetry is broken and the sneutrinos acquire vacuum expectation values, $\mu=\lambda_i \langle \tilde \nu^c_i\rangle$. 
On the other hand, mixing terms among right-handed neutrinos, $\kappa_{ijk}\hat \nu^c_i \hat \nu^c_j \hat \nu^c_k$,
contribute to generate effective Majorana masses for neutrinos at the electroweak scale, $\sim
\kappa_{ijk} \langle \tilde \nu^c_k \rangle$. With neutrino Yukawa couplings $Y_{\nu} \lesssim 10^{-6}$, this dynamically
generated electroweak-scale seesaw mechanism can easily reproduce current
measurements of neutrino mass differences and mixing angles~\cite{Escudero:2008jg,neutrinos,Bartl:2009an,neutrinos0,neutrinos2}.
Given that the $\mu\nu$SSM is a well motivated and attractive model, its phenomenology at the LHC has been analysed in several works~\cite{fidalgo,Bartl:2009an,gh,gh2,neutrinos,neutrinos2,pradipta,Ghosh:2014rha}. Cosmological issues in this model have also been considered, and in particular the generation of the baryon asymmetry of the Universe was studied in detail~\cite{chung,valle}, with the interesting result that electroweak baryogenesis can be realised~\cite{chung} while thermal leptogenesis is disfavoured in the context of the $\mu\nu$SSM~\cite{valle}. 

The mixing terms characterising the $\mu\nu$SSM produce an explicit breaking of 
$R$-parity. The size of the breaking is small, since it is determined by $Y_{\nu}$. As a consequence of the
$R$-parity violation, the lightest supersymmetric particle (LSP) is no longer
stable. Thus, the lightest neutralino/sneutrino would have a very short lifetime and could not be a viable DM candidate. 
Nevertheless, if the role of the LSP is played by the gravitino, its decay is suppressed both by the
feebleness of the gravitational interaction and by the small $R$-parity
violating coupling, and, as a consequence, its lifetime can be much longer than the
age of the Universe. In addition, the gravitino can be produced by thermal scatterings in the early Universe 
with a relic density matching the observed DM density in the Universe.
Thus, the gravitino, which is a superweakly interacting massive particle (superWIMP), represents a good DM candidate. Most importantly, as pointed out in ref.~\cite{yamaguchi} for the case of $R$-parity
violation, gravitino decays in the Milky Way halo
would produce monochromatic \gammaRays\ with an energy equal to half of the
gravitino mass, and therefore its presence can, in principle, be inferred
indirectly from \gammaRayHyph\ observations.\footnote{The \gammaRayHyph\ line
signature from gravitino DM decay is not an exclusive feature of the
$\mu\nu$SSM. Other gravitino DM scenarios with bilinear or trilinear $R$-parity
violation as discussed in refs.~\cite{Buchmuller:2007ui,Lola:2007rw,Diaz:2011pc,taoso} may
exhibit the same decay signature~\cite{Ibarra:2007wg}. In addition, this
signature also appears in models with axino DM decay via bilinear $R$-parity
violation~\cite{Kim:2001sh,Covi:2009pq}. The \gammaRayHyph\ line constraints
derived in this work thus could also be applied to constrain those models.}

Several searches for DM-induced \gammaRayHyph\ lines have been performed using
\Fermi\/-LAT data.  One of the first explicit searches for \gammaRayHyph\ lines from gravitino DM in the $\mu\nu$SSM was performed in~\cite{Choi:2009ng}. From the non-observation of prominent sharp features in the diffuse emission measurement (based on 5 months of data) reported by the \Fermi\/-LAT collaboration~\cite{Abdo:2009mr}, gravitinos with masses larger than 10\,GeV turn out to be disfavoured, as well as lifetimes smaller than about 3 to $5\times 10^{27}$\,s. In ref.~\cite{Vertongen:2011mu}, 2 years of \Fermi\/-LAT data were used to constrain  \gammaRayHyph\ lines in the energy range between 1\,GeV and 300\,GeV. Stringent lower bounds on the gravitino lifetime of about $5\times 10^{28}$\,s were obtained for masses above 2\,GeV. When these bounds are applied to the $\mu\nu$SSM, they imply that the gravitino mass must be smaller than 4\,GeV~\cite{gustavo}. At somewhat lower energies, limits have also been established from observations of the Galactic Centre by the 
Energetic Gamma Ray Experiment Telescope (EGRET) on board the {\it Compton Gamma Ray Observatory}~\cite{EGRET}. 
In refs.~\cite{Abdo:2010nc,Bloom,Fermi-LAT:2013uma}, the \Fermi\/-LAT collaboration presented constraints on monochromatic \gammaRayHyph\  emission. In particular, in ref.~\cite{Fermi-LAT:2013uma} using 44 months of data, the derived limits refer to the emission above 5\,GeV, covering, in the context of gravitino DM, masses larger than 10\,GeV. As we will show, this limit implies that gravitinos with masses larger than 10\,GeV are excluded in the $\mu\nu$SSM.

Given these previous results, and the interest of the $\mu\nu$SSM as an
attractive supersymmetric model that will be tested at the LHC, an extension of
the analyses, covering line energies below a few GeV, is of great importance.
In this work we report on a search for \gammaRayHyph\ spectral lines from 100\,MeV to 10\,GeV using 62 months of \Fermi\/-LAT data. In this energy range, because of the small
statistical uncertainties, the analysis is dominated by
systematic effects that may fake or mask a line signal.
Therefore, for the first time, we present \gammaRayHyph\ line limits where
systematic uncertainties are included in the likelihood fitting and the
calculation of limits. 

This work is organised as follows. In section~\ref{munussm}, gravitino DM in
the $\mu\nu$SSM is introduced, paying special attention to its lifetime and associated relic density. In section~\ref{data}, after discussing the DM distribution we adopted as a baseline for this analysis, we concentrate on the data analysis and our treatment of systematic uncertainties
and derive constraints on the parameter space of both generic decaying
and annihilating DM. Finally, in section~\ref{sec:discussions} the constraints on
decaying DM are applied to the $\mu\nu$SSM gravitino DM model and our results are
compared with previous limits reported in the literature. The conclusions are
left for section~\ref{conclusions}.

\section{\texorpdfstring{Gravitino dark matter in the \boldmath$\mu\nu$SSM}{Gravitino dark matter in the munuSSM}} \label{munussm}
\subsection{Gravitino lifetime}

In the supergravity Lagrangian there is an interaction term between the gravitino, $\Psi_{3/2}$, the field strength for the photon, and the photino. Due to the breaking of $R$-parity, the photino and the left-handed neutrinos are mixed, and thus the gravitino will be able to decay, through the interaction term, into a photon and a neutrino with energies equal to half of the gravitino mass, $m_{3/2}$~\cite{yamaguchi}.\footnote{The gravitino could also decay into a $W^{\pm}$ and a charged lepton, into a $Z^0$ and a neutrino, or into a Higgs boson and a neutrino~\cite{Ishiwata:2008cu}. However, these decay channels are not kinematically accessible in our case, since the mass of the gravitino in the $\mu\nu$SSM must be smaller than about 10\,GeV in order to fulfil the observational constraints.
Also, because of this upper bound on the gravitino mass, three-body decay modes of the gravitino~\cite{Choi:2010jt,aurelio,Grefe:2011dp} are not relevant, and we will not consider them throughout this work.} Therefore, the presence of the gravitino can, in principle, be inferred indirectly from observations of the diffuse backgrounds of photons or neutrinos.\footnote{The flux of monochromatic neutrinos could be in principle observed in neutrino detectors. However, at energies around 1\,GeV the signal is expected to be overwhelmed by atmospheric neutrinos and, given the typically bad neutrino energy resolution, also a spectral analysis is not of much help. Moreover, the effective volume of neutrino detectors in the GeV range is too small to expect a sizeable signal event rate. See ref.~\cite{Covi:2008jy} for a related discussion. Thus we concentrate on \gammaRayHyph\  line searches throughout this work.}

The lifetime of the gravitino LSP in the $\mu\nu$SSM is typically much longer than the age of the Universe, making it a viable DM candidate. From the supergravity Lagrangian one obtains
a decay width, $\Gamma$, given by~\cite{yamaguchi}%:
\begin{linenomath}
\begin{equation}
\Gamma\left(\Psi_{3/2}
\to \sum_i \gamma \nu_i\right)\simeq  \frac{1}{32 \pi}\, |U_{\widetilde{\gamma} \nu}|^2\,\frac{m^3_{3/2}}{M_P^2}\,, 
\label{c1}
\end{equation}
\end{linenomath}
where $M_P=2.4\times 10^{18}$\,GeV is the reduced Planck
mass, and $|U_{\widetilde{\gamma}\nu}|^{2}$ determines the neutrino
content of the photino:
\begin{linenomath}
\begin{equation}
|U_{\widetilde{\gamma}\nu}|^{2}=\sum_{i=1}^{3}|N_{i1}\cos\theta_{W}+N_{i2}\sin\theta_{W}|^{2}.
\end{equation}
\end{linenomath}
Here $N_{i1}$ ($N_{i2}$) is the Bino (Wino) component of the $i$-th neutrino, and $\theta_{W}$ is the weak mixing angle.  
The gravitino lifetime can then be written as: 
\begin{linenomath}
\begin{equation}
  \tau_{3/2}\simeq 3.8\times10^{27}\,\text{s}\left(\frac{|U_{\widetilde{\gamma}\nu}|^2}{10^{-16}}\right)^{-1}\left(\frac{m_{3/2}}{10\,\text{GeV}}\right)^{-3}.
\label{munuSSMlifetime}
\end{equation}
\end{linenomath}

Since the electroweak-scale seesaw mechanism that is needed to reproduce neutrino data ~\cite{Tortola:2012te,GonzalezGarcia:2012sz} in the $\mu\nu$SSM is determined by the neutrino Yukawa couplings $Y_{\nu} \lesssim 10^{-6}$, this dictates a very small mixing between the photino and the neutrinos giving rise to the approximate range for $|U_{\widetilde{\gamma}\nu}|^{2}$~\cite{Choi:2009ng}:
\begin{linenomath}
\begin{equation}
10^{-16} \lesssim |U_{\widetilde{\gamma}\nu}|^{2} \lesssim 10^{-12}.
\label{representative}
\end{equation}
\end{linenomath}
Taking into account eq.~(\ref{munuSSMlifetime}), this estimate implies that the gravitino will be very long lived compared to the
current age of the Universe, which is about $4\times 10^{17}$\,s. As discussed in~\cite{Choi:2009ng}, one can carry out the numerical analysis of the whole low-energy parameter
space of the $\mu\nu$SSM, confirming the result~(\ref{representative}). Nevertheless, these bounds are very
conservative, and in fact the results of the numerical scan in~\cite{Choi:2009ng} favour the much smaller range:
\begin{linenomath}
\begin{equation}
  10^{-15}\lesssim|U_{\widetilde{\gamma}\nu}|^2\lesssim5\times10^{-14}.
\label{scan}  
\end{equation}
\end{linenomath}
It is worth noting that in the scan of ref.~\cite{Choi:2009ng} the mass of the lightest
neutralino is typically above 20\,GeV, and since $m_{3/2}$ is constrained to be smaller
than a few GeV in the $\mu\nu$SSM, as discussed in the introduction, the gravitino can safely be
used as the LSP. 
The ranges for the photino--neutrino mixing found in eqs.~(\ref{representative}) and~(\ref{scan}) together with the formula for the gravitino lifetime, eq.~(\ref{munuSSMlifetime}), give a clear prediction for the expected \gammaRayHyph\ line signal from gravitino decays in the $\mu\nu$SSM that will be tested against the \Fermi\/-LAT data in our analysis below.

Let us finally remark that for the gravitino to be a good DM candidate
also requires that it can be present in the right amount to explain
the relic density inferred by cosmological observations, $\Omega_{\text{DM}}
h^2 \simeq 0.1$. We will discuss this issue in the next subsection.

\subsection{Gravitino relic density}\label{cosmology}

An inflationary phase in the early
Universe, as supported by many cosmological observations, would dilute any
primordial abundance of gravitinos. In many cases, depending on the values of the reheating
temperature $T_R$ and $m_{3/2}$, the gravitino would not reach
thermal equilibrium with the rest of the hot plasma after
inflation~\cite{Ellis:1982yb}.  Still, gravitinos could be produced in scattering
processes in the thermal bath. The relic density of gravitinos from thermal
production (TP) in the early Universe is given by~\cite{Pradler:2006qh}:
\begin{linenomath}
\begin{equation} \Omega_{3/2}^{\text{TP}}h^2\simeq\sum_{i=1}^3\omega_i\,g_i^2\left(1+\frac{M_i^2}{3\,m_{3/2}^2}\right)\ln\left(\frac{k_i}{g_i}\right)\left(\frac{m_{3/2}}{100\,\text{GeV}}\right)\left(\frac{T_R}{10^{10}\,\text{GeV}}\right),
\label{gravitinorelic1}
\end{equation}
\end{linenomath}
where the sum runs over the SM gauge groups. The numerical factors are given by $\omega_i=(0.018, 0.044, 0.117)$ and 
$k_i=(1.266, 1.312, 1.271)$. The gauge couplings $g_i$ and the gaugino masses $M_i$ are understood to be evaluated at an 
energy corresponding to the reheating temperature. The one-loop renormalisation group equations for these parameters are 
given by~\cite{Pradler:2006hh}:
\begin{linenomath}
\begin{align}  g_i(T_R)&=\left[g_i(m_Z)^{-2}-\frac{\beta_i^{\text{SM}}}{8\,\pi^2}\ln\left(\frac{m_{\text{SUSY}}}{m_Z}\right)-\frac{\beta_i^{\text{SUSY}}}{8\,\pi^2}\ln\left(\frac{T_R}{m_{\text{SUSY}}}\right)\right]^{-1/2},\\
  M_i(T_R)&=\left(\frac{g_i(T_R)}{g_i(m_Z)}\right)^2M_i(m_Z)\,,
\end{align}
\end{linenomath}
where the SM beta function coefficients are $\beta_i^{\text{SM}}=(41/6, -19/6, -7)$, the MSSM beta 
function coefficients are $\beta_i^{\text{SUSY}}=(11, 1, -3)$\footnote{As the $\mu\nu$SSM only extends the particle content 
compared to the MSSM by three right-handed neutrinos, the $\beta$ function coefficients do not change.} and the values for the 
gauge couplings at the electroweak scale are $g_2(m_Z)\equiv g(m_Z)=m_W\sqrt{8\,G_F/\smash{\sqrt{\vphantom{a}\smash{2}}}}\simeq0.65$, $g_1(m_Z)\equiv g'(m_Z)=g_2(m_Z)\tan{\theta_W}(m_Z)\simeq0.36$ and $g_3(m_Z)\equiv g_s(m_Z)=\sqrt{4\,\pi\,
\alpha_s(m_Z)}\simeq1.22$~\cite{Beringer:1900zz}. In these expressions $m_W$ and $m_Z$ are the masses of the $W$ and $Z$ bosons, respectively, and $G_F$ is the Fermi constant. We assume a supersymmetry mass scale $m_{\text{SUSY}}\sim 1\,$TeV.

Equation~(\ref{gravitinorelic1}) only holds for
gravitino densities well below the thermal equilibrium density since only
gravitino production processes and no reverse processes are taken into account
in the derivation~\cite{Bolz:2000fu,Pradler:2006qh}. That is, in cases where we
want to fix the gravitino abundance to the measured DM density in the Universe,
the formulae are valid for gravitino masses above
$\mathcal{O}(1)\,$keV~\cite{Pagels:1981ke}. 
Moreover, the derivation makes
use of an expansion in the coupling constants that is only well justified for
temperatures $T_R\gg10^6\,$GeV~\cite{Bolz:2000fu,Pradler:2006qh}. 
Thus the extrapolation to lower reheating temperatures may not be entirely reliable, 
but it is the best estimate we have at this time.

% created this plot with MathematicaPlots.nb
\begin{figure}[t]
  \centering
  \includegraphics[width=0.8\twocolfigwidth]{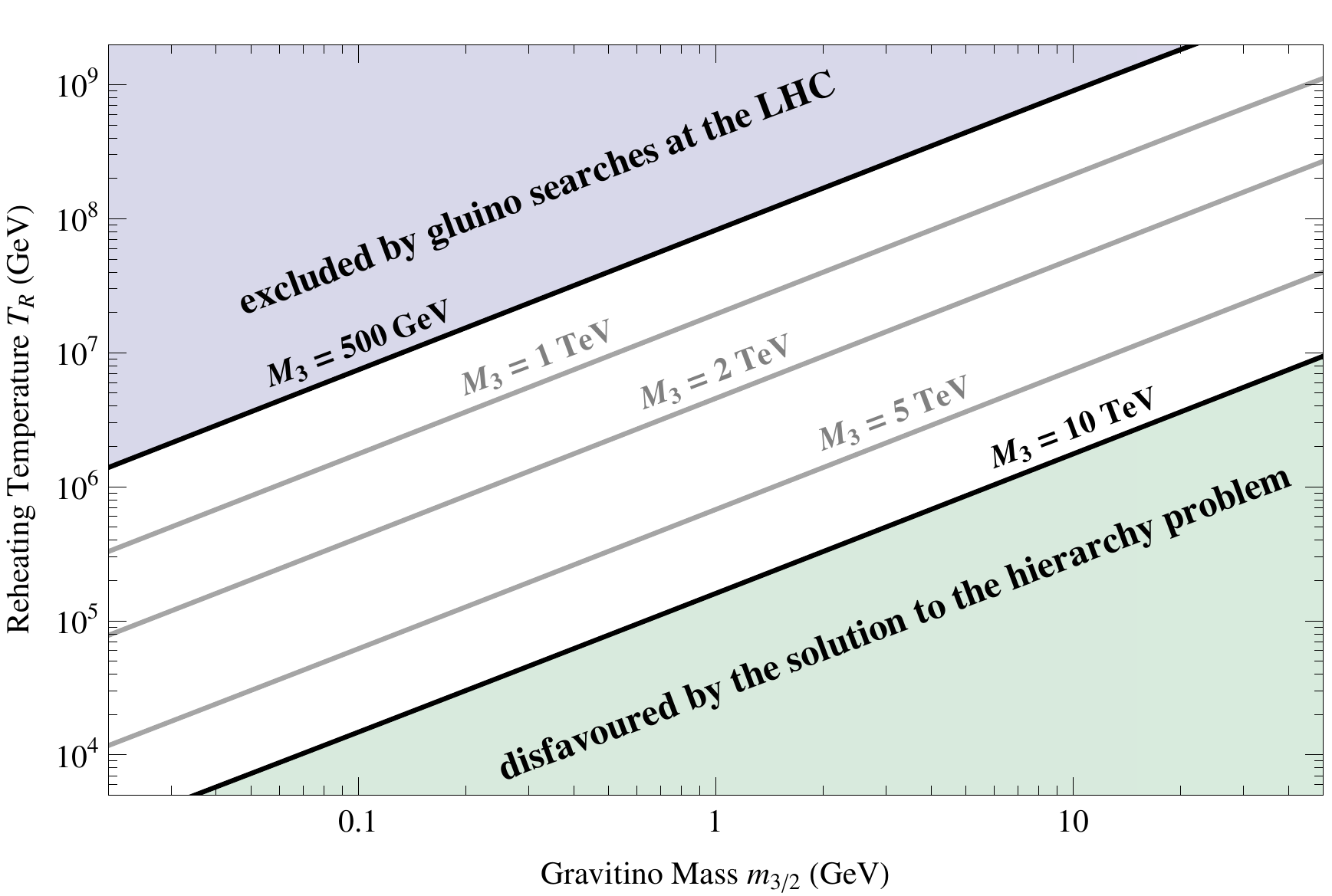}
  \caption{Parameter space for the gravitino DM scenario in the
      $m_{3/2}$--$T_R$ plane. The diagonal lines show the contours where the gravitino relic density matches the
  observed DM density for the indicated value of the gluino mass according to
  eq.~\eqref{gravitinorelic1}. We assume universal gaugino masses at the
  GUT scale. The barely visible widths of the lines correspond to the $3\sigma$ uncertainty of the cosmological data. The areas above the $M_3=500\,$GeV line and below the $M_3=10\,$TeV line are excluded.}
  \label{fig:relicdensity}
\end{figure}

In figure~\ref{fig:relicdensity} we show the gravitino mass--reheating
temperature plane. If we require the gravitino abundance to match the abundance
of cold DM in the Universe as determined by a combination of different cosmological
observables, $\Omega_{\text{DM}}h^2=0.1187\pm0.0017$~\cite{Ade:2013zuv}, 
fixed values for the gaugino masses correspond to lines in this plane. The widths of those lines 
correspond to the $3\sigma$ uncertainty of the cosmological data, thus showing the high level of accuracy in the experimental determination of the DM relic density.
We show here the situation for the case of universal gaugino masses at the GUT scale, corresponding to mass ratios of $M_3\simeq2.8\,M_2\simeq5.1\,M_1$ at the scale $m_{\text{SUSY}}\sim 1\,$TeV. Giving up this relation will change the picture but the results remain similar at least as long as the gluino is the heaviest of the gauginos. The ATLAS and CMS collaborations at the LHC have performed several searches for signals of supersymmetry, so far without success. Depending on the model assumptions they find lower limits on the gluino mass $M_3$ ranging from several hundred GeV to beyond a TeV~\cite{ATLAS_SUSY,CMS_SUSY}. 
There are currently no gluino mass limits by the LHC collaborations for the specific case of the $\mu\nu$SSM or models with gravitino DM and $R$-parity violation in general.
Thus, for definiteness we use in figure~\ref{fig:relicdensity} a conservative lower limit of 500\,GeV. On the other hand, we expect a gluino mass below $\mathcal{O}(10)\,$TeV to solve the hierarchy problem by means of supersymmetry. 
This leaves us with the white area as the viable parameter space for gravitino DM.

As we will discuss below, strong bounds on the parameter space of decaying
gravitino DM can be derived from \gammaRayHyph\ line searches. 
The result of our analysis allows us to limit the gravitino mass to values below 4.8\,GeV in the case of the $\mu\nu$SSM 
(see section~\ref{results}).\footnote{Since the mechanism of supersymmetry breaking is unknown, in supergravity models like 
the $\mu\nu$SSM the gravitino mass is basically a free parameter. However, one can apply observational constraints like 
the lower limit of 1.2\,keV on the mass of warm DM particles derived from Lyman-alpha forest 
data~\cite{Viel:2005qj,Viel:2007mv}.}
This result, in turn, restricts the reheating temperature to be below $\sim4\times10^8\,$GeV (cf.\ 
figure~\ref{fig:relicdensity}).\footnote{Since there is no unique theory of inflation and the reheating phase of the 
early Universe, there are almost no theoretical constraints on the value of the reheating temperature. From observational
 constraints the reheating temperature could be as low as a few MeV~\cite{Hannestad:2004px}.}
As mentioned in the introduction, it has been found that in the context of the $\mu\nu$SSM the baryon asymmetry of the Universe can be generated by the mechanism of electroweak baryogenesis which does not require a very high value of the reheating temperature~\cite{chung}. Other possibilities, like the popular mechanism of thermal leptogenesis~\cite{Fukugita:1986hr}, are in tension with the upper limit on the reheating temperature found above. However, leptogenesis is disfavoured anyway in the context of the $\mu\nu$SSM since the model exhibits an electroweak-scale seesaw mechanism. 

Apart from producing the correct relic density, common problems in supergravity theories are conflicts with the predictions of big bang nucleosynthesis (BBN) due to late decays of the next-to-lightest supersymmetric particle (NLSP) into the gravitino~\cite{Kawasaki:2008qe}. In the $\mu\nu$SSM this problem is easily evaded since the NLSP decays well before the onset of BBN via $R$-parity breaking interactions (see ref.~\cite{Buchmuller:2007ui} for a related discussion). 

Let us conclude this section with the remark that there is plenty of parameter
space for a gravitino LSP with the correct relic density and 
leading to a consistent cosmological scenario fulfilling all current constraints.

\section{\texorpdfstring{Search for \boldmath\gammaRayHyph\ lines from below 10\,GeV in the \Fermi\/-LAT
data}{Search for gamma-ray lines from below 10 GeV in the \Fermi\/-LAT data}}
\label{data}

\subsection{Gamma-ray flux and dark matter distribution}

Cosmological N-body simulations predict the inner part of the Galaxy to enclose the highest DM density and have inspired parametrisations of the DM halo distribution.  
We calculate the differential flux of $\gamma$ rays from DM decays in the
Galactic halo by integrating one particular DM distribution along the line of sight.  We assume here two-body decays producing monochromatic $\gamma$ rays and neutrinos. In this case the flux reads:
\begin{linenomath}
\begin{equation}\label{eq:decayFlux}
 \frac{d\Phi_{\gamma,\,\text{dec}}^{\text{halo}}}{dE}=\frac{1}{4\,\pi\,\tau_{\gamma\nu}\,m_{\text{DM}}}\,\delta\left(E-\frac{m_{\text{DM}}}{2}\right)\int_{\Delta\Omega}\!\!\cos
 b\,db\,d\ell\int_0^{\infty}\!\! ds\,\rho_{\text{halo}}(r(s,\,b,\,\ell))\,,
\end{equation}
\end{linenomath}
where $b$ and $\ell$ denote the Galactic latitude and longitude,
respectively, and $s$ denotes the distance from the Solar System.  Furthermore,
$m_{\text{DM}}$ and $\tau_{\gamma\nu}$ are the DM mass and lifetime (here inverse
decay width for $\text{DM}\to\gamma\nu$), respectively, $\Delta \Omega$ is the region of interest (ROI), 
and
$\delta$ denotes the Dirac delta distribution.
The radius $r$ in the DM halo density profile of the Milky Way, $\rho_{\text{halo}}$, is expressed in terms of these Galactic coordinates,
\begin{linenomath}
\begin{equation}
 r(s,\,b,\,\ell)=\sqrt{s^2+R_{\odot}^2-2\,s\,R_{\odot}\cos{b}\cos{\ell}}\,.
\end{equation}
\end{linenomath}
In this expression $R_{\odot}\simeq8.5\,\text{kpc}$~\cite{Ghez:2008ms,Gillessen:2008qv,Catena:2009mf}
is the radius of the solar orbit around the Galactic Centre. The corresponding
formula for the \gammaRayHyph\ flux from self-conjugate DM annihilations via $\text{DM DM}
\to \gamma\gamma$ reads:
\begin{linenomath}
\begin{equation}\label{eq:annihilationFlux}
 \frac{d\Phi_{\gamma,\,\text{ann}}^{\text{halo}}}{dE}=\frac{\left\langle\sigma
 v\right\rangle_{\gamma\gamma}}{8\,\pi\,m_{\text{DM}}^2}\,2\,\delta\left(E-m_{\text{DM}}\right)\int_{\Delta\Omega}\!\!\cos
 b\,db\,d\ell\int_0^{\infty}\!\! ds\,\rho_{\text{halo}}^2(r(s,\,b,\,\ell))\,,
\end{equation}
\end{linenomath}
where $\langle\sigma v\rangle_{\gamma\gamma}$ is the thermal-averaged DM annihilation cross section into two photons. 
For the later discussion it is useful to define $J$-factors, which are directly
proportional to the relevant line-of-sight integral and the signal flux in a
given ROI that spans the solid angle $\Delta\Omega_{\text{ROI}}$,
\begin{linenomath}
\begin{align}
    J_{\text{dec\,(ann)}} = \int_{\Delta\Omega_{\text{ROI}}}
    \!\!\!\!\!\!\!\!\!\cos b\,db\,d\ell \int_0^\infty \!\! ds\ \rho_{\text{halo}}^{1(2)} (r(s, b,
    \ell))\;.
\end{align}
\end{linenomath}
Throughout this analysis we will employ the Einasto profile with a finite central density~\cite{1965TrAlm...5...87E,Navarro:2003ew}:
\begin{linenomath}
\begin{equation}
 \rho_{\text{Ein}}(r)=\rho_{\odot}\exp\left( -\frac{2}{\alpha}\left( \left( \frac{r}{r_s}\right) ^\alpha-\left( \frac{R_{\odot}}{r_s}\right) ^\alpha\right) \right) ,
\end{equation}
\end{linenomath}
where we adopt $\alpha=0.17$ and $r_s=20\,$kpc for the case of the Milky Way and a local DM density of
$\rho_{\odot}\simeq0.4\,\text{GeV}\,\text{cm}^{-3}$~\cite{Catena:2009mf,Weber:2009pt,Salucci:2010qr}.
In appendix~\ref{app:halo_profile} we consider other DM halo profiles as well as uncertainties on the 
halo parameters and quantify the impact of the choices on our results.

\subsection{Selection and processing of \Fermi\/-LAT data}
We search for \gammaRayHyph\ spectral lines from 100\,MeV to
10\,GeV.  To include the spectral sidebands for all fit points, we consider data from 56.5\,MeV to 11.5\,GeV.  For our dataset we use the \irf{P7REP\_CLEAN}\ event selection on data taken between August 4, 2008, and October 15, 2013 by the \Fermi\/-LAT.  We chose to use the stringent \irf{P7REP\_CLEAN}\ event selection since it has low residual cosmic-ray contamination compared to the \gammaRayHyph\ flux.  More information about the \Fermi\/-LAT instrument, performance, and  data usage can be found in refs.~\cite{REF:2009.LATPaper,REF:2012.P7Perf} as well as the FSSC website.\footnote{The \Fermi\/-LAT photon data are available at \url{http://fermi.gsfc.nasa.gov/ssc/}.} A short overview of the \Fermi\/-LAT instrument and event class naming convention can be found in section~II of ref.~\cite{Fermi-LAT:2013uma}.

The version of the instrument response functions (IRFs) that we used in this work is
\texttt{P7REP\_CLEAN\_V15}. We only use events with a measured zenith angle less than 100$^{\circ}$ to remove the emission from the Earth's
limb (i.e.\ \gammaRays\ from cosmic-ray interactions in the upper atmosphere).  We also apply the standard
good time selection criteria "\texttt{DATA\_QUAL == 1 \&\& LAT\_CONFIG == 1 \&\& abs(ROCK\_ANGLE) < 52}" using the
\gtmktime\ \texttt{ScienceTool}.  Note that the adopted rocking angle cut is only
applicable to data taken prior to December 6, 2013, which is the case for our dataset.
The initial data reduction and exposure calculations were performed using the
\Fermi\/-LAT \texttt{ScienceTools} version 09-32-02.\footnote{\url{http://fermi.gsfc.nasa.gov/ssc/data/analysis/software/}} 
Further details about the data selection cuts in signal and control regions can be found in
appendix~\ref{app:control_samples}.
 
We have not performed any point source masking and do not include point sources in our fitting
procedure.  We choose not to mask sources because at the low end of
the energy range of our dataset, the $68\%$ containment radius of the
\Fermi\/-LAT point-spread function (PSF) is
$\gtrsim5^{\circ}$~\cite{REF:2012.P7Perf}, which would cause us to mask almost the entire dataset given the large number of known
\gammaRayHyph\ point sources.  However, \gammaRayHyph\ point sources are not
expected to produce narrow line-like spectral features; we will quantify the
systematic uncertainties that follow from our analysis choices in
section~\ref{sec:syst}.

\subsection{Region of interest optimisation}\label{subsec:ROI}
Since a DM signal would have a very different morphology from the dominant astrophysical \gammaRayHyph\ emission, optimising the ROIs used is critical for an efficient search for \gammaRayHyph\ lines~\cite{Bringmann:2012vr,
Weniger:2012tx}. In cases where the search is background limited and
systematics can be neglected, the statistical power of a line search is
maximised if the signal-to-noise ratio, $n_\text{sig}/\sqrt{b_\text{eff}+n_\text{sig}}$, within the ROI is
maximal.  Here, $n_\text{sig}$ denotes the number of expected signal events, $b_\text{eff}$ the
number of effective background events (for a definition see
Eq.~\eqref{eqn:beff} below), and a useful common assumption is that
$n_\text{sig}\ll b_\text{eff}$. An optimisation with respect to this signal-to-noise ratio is relevant
in searches for \gammaRayHyph\  lines at intermediate and high \Fermi\/-LAT
energies.
However, as we will discuss below in more detail, at energies $\lesssim3$\,GeV the search will be limited by systematic effects that
scale approximately linearly with the number of background \gammaRays.  In that case, an optimal
ROI should rather maximise the signal-to-background ratio,
$n_\text{sig}/b_\text{eff}$, in order to achieve maximal discrimination power between a real
signal, an instrumental effect and background modelling uncertainties.

\begin{figure}[t]
  \centering
  \includegraphics[width=0.8\twocolfigwidth]{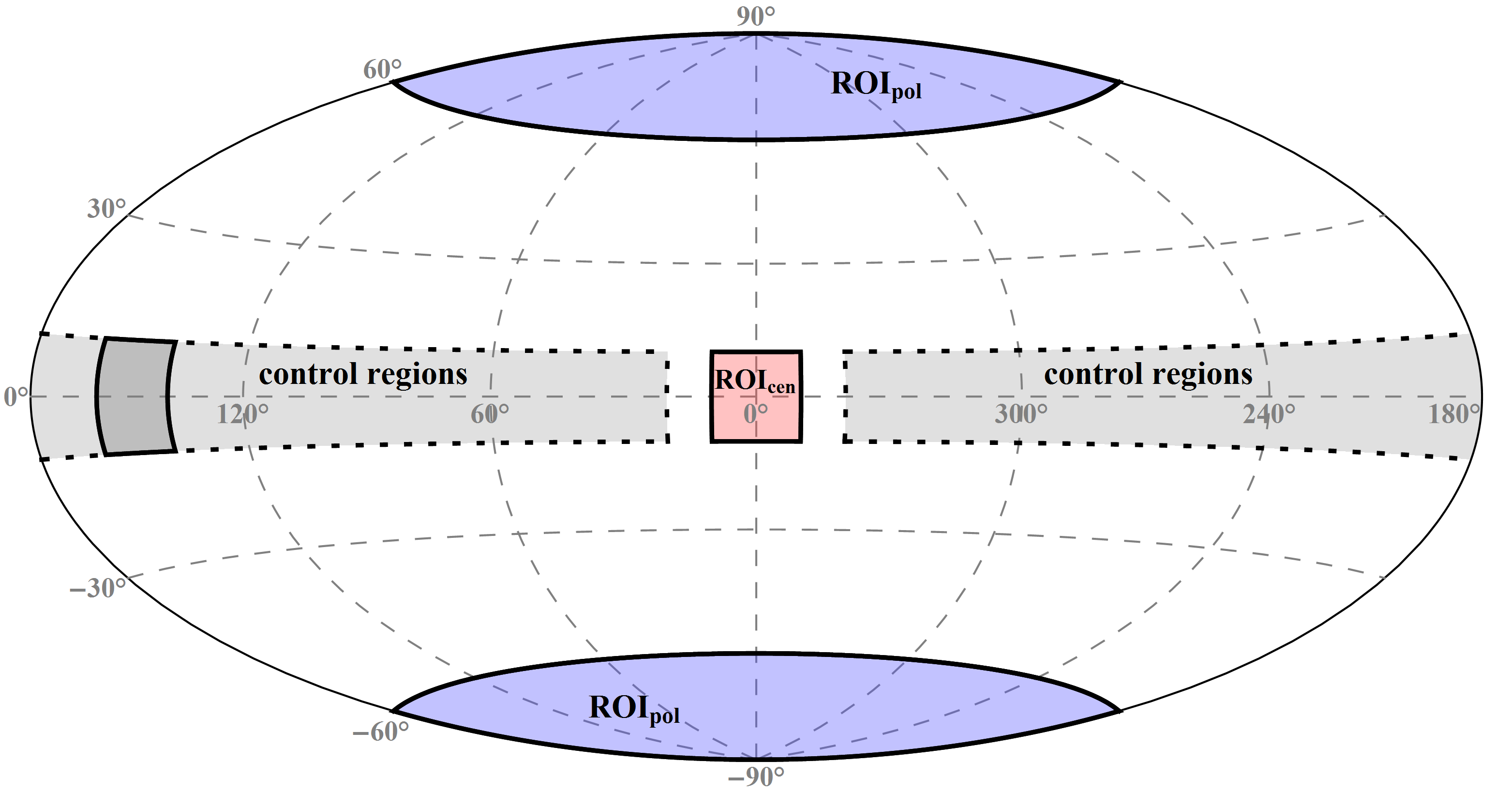}
  \caption{Skymap of ROIs used in this analysis; plotted in Galactic coordinates using the Hammer--Aitoff projection. The region ROI$_{\text{pol}}$ (blue) is optimised for the signal-to-background ratio in the case of DM decay, while the region ROI$_{\text{cen}}$ (red) is optimised for the signal-to-background ratio in the case of DM annihilation. The dashed line encloses the area for the control regions along the Galactic plane (light grey), while the grey region is an example of one of the 31 control regions used in this analysis.}
  \label{fig:rois}
\end{figure}

In the present analysis, we select two ROIs that optimise the
signal-to-background ratio for searches for
decay and annihilation signals, respectively.  The details of the ROI optimisation
process are in appendix~\ref{app:ROIs}.  For our analysis, we selected
the following two ROIs:
\begin{linenomath}
\begin{align}
    &\text{ROI$_{\text{pol}}$:}\; |b|>60^\circ & &\text{for decay, and}  \\
    &\text{ROI$_{\text{cen}}$:}\; (|b|<10^\circ \;\;\text{and} \;\; |\ell|<10^\circ) \hspace{-20mm}& &\text{for annihilation,}
\end{align}
\end{linenomath}
where the decay ROI includes the Galactic poles and the annihilation ROI includes the Galactic Centre.
Note that the size of the search region ROI$_{\text{cen}}$ is chosen in order to be larger than the PSF at low energies (the 68\% containment at 100\,MeV is $\sim 5^{\circ}$). The signal-to-background ratio for the ROI$_{\text{cen}}$ region is roughly a factor of two less than the optimal value would be if the PSF could be neglected. 
The positions and sizes of these regions are indicated in 
figure~\ref{fig:rois}. We calculated the $J$-factors corresponding to an
annihilation or decay signal for both regions. In case of the Einasto profile,
they are $J=2.89\times10^{22}\,\text{GeV\,cm}^{-2}$ for decay and
$J=8.89\times10^{22}\,\text{GeV}^2\,\text{cm}^{-5}$ for annihilation.   The
$J$-factors for other DM profiles are given in appendix~\ref{app:halo_profile}.

% created this plot with MathematicaPlots.nb

\subsection{Fitting procedure}\label{sec:FitProc}

Our search for a line signal in the \Fermi\/-LAT data, as well as the derivation
of upper limits on line fluxes, is based on the profile likelihood method (see
e.g.\ ref.~\cite{Rolke:2004mj}).  We model the  sum of the astrophysical \gammaRayHyph\ background and the cosmic-ray contamination of the \irf{P7REP\_CLEAN}\ data empirically by a single power law with free
normalisation and spectral index.  Since the power-law approximation is only
valid locally and breaks down when considering large enough energy ranges,
we restrict the fit to small energy ranges centred around and moving with the
line energy.  In the present work, we adopt an energy range of
$(E_\gamma - 2\sigma_E, E_\gamma+ 2\sigma_E)$, where $E_\gamma$ denotes the line
energy of interest, and $\sigma_E/E_\gamma$ is the energy resolution at that energy
($\pm\sigma_E$ is the 68\% containment range).  We selected this energy range as
a compromise between a loss of statistical power in smaller ranges, and
increasing systematic uncertainties in cases of larger ranges. Each fit was performed at a fixed energy, $E_\gamma$, in steps of $0.5\sigma_E$, where $\sigma_E$ ranges from 20\% of $E_\gamma$ at 100\,MeV to 10\% of $E_\gamma$ at 10\,GeV.  We used the \texttt{RooFit} toolkit~\cite{REF:RooFit.2003} (version 3.12) to implement the models and perform the likelihood minimisation.
\medskip

At the low energies of interest, the number of photon events in our analysis
is very large.  For computational efficiency, we perform a binned
maximum likelihood fit to the data, with a bin width of $0.066 \sigma_E$ (i.e.\ 60 bins over the
$\pm2\sigma_E$ energy window).  
Furthermore, we take into account the possibility that the true number of
signal events, $n_{\text{sig}}$, is systematically offset by $n_{\text{syst}}$ from the best fit value,
$n'_{\text{sig}}$.  In other words, we only consider the true signal events to
be those that remain after subtracting the expected systematic offset,
$n_{\text{sig}} = n'_{\text{sig}} - n_{\text{syst}}$,
and taking into account its variance.
The full likelihood function that we adopt in our analysis is based on the product of the Poisson likelihoods ($P$) to observe $c_i$ counts in each energy bin:
\begin{linenomath}
\begin{equation}
    \mathcal{L}(\alpha, \Gamma, n_{\text{sig}}, n_\text{syst}) =
    P_\mathcal{F}(n_{\text{syst}}, b_\text{eff}) \prod_i
    P(c_i| \mu_i(\alpha, \Gamma, n_{\text{sig}} + n_{\text{syst}}%(\delta f_{\text{syst}},b_{\text{eff}})
    ) )\;,
    \label{eqn:likelihood}
\end{equation}
\end{linenomath}
where the expected number of events in the $i$-th energy bin $(E_i^- \equiv E_{\gamma}-2\sigma_E, E_i^+ \equiv E_{\gamma}+2\sigma_E)$ is
given by
\begin{linenomath}
\begin{equation}
    \mu_i(\alpha, \Gamma, n'_{\text{sig}}) = \int_{E_i^-}^{E_i^+} dE \left( \alpha
    E^{-\Gamma}\mathcal{E}(E) + n'_{\text{sig}}\cdot \mathcal{D}_{\text{eff}}(E|E_\gamma) \right)\;,
\end{equation}
\end{linenomath}
and $\mathcal{D}_{\text{eff}}(E|E_\gamma)$ denotes the energy dispersion
of the \Fermi\/-LAT. Furthermore, $\mathcal{E}(E)$ denotes the energy-dependent
exposure of the ROI, $\Gamma$ and $\alpha$ are the spectral index and
normalisation of the power-law background, $b_{\text{eff}}$ is the effective
number of background events in the energy range covered by the line signal, $n_{\text{syst}}$ is the
additive systematic error (to be discussed below), and $P_\mathcal{F}$ is the
distribution of $n_{\text{syst}}$, which we model to be independent of energy.  Note that we actually fitted for
$n_{\text{bkg}} = \int \alpha E^{-\Gamma}\mathcal{E}(E)  dE$, the
total number of events in the power-law background, rather than $\alpha$
directly.
\medskip

As discussed in appendix~C5 of ref.~\cite{Fermi-LAT:2013uma}, $\mathcal{D}_{\text{eff}}$ 
varies slightly depending on the ``observing profile'' (i.e.\ the
amount of observing time for each event incident angle, $\theta$).  To account for this
in our search, we modelled $\mathcal{D}_{\text{eff}}$ for each fit similar to what
was done in ref.~\cite{Weniger:2012tx}. Specifically, we integrated
the energy- and $\theta$-dependent representation of the energy
dispersion provided with the \Fermi\/-LAT IRFs over the observing profile for
the regions of interest and then fit a triple Gaussian (sum of three Gaussian
functions) parametrisation
to that shape to serve as our $\mathcal{D}_{\text{eff}}$ model. 

We furthermore investigated the effect of including additional information in
our $\mathcal{D}_{\text{eff}}$ model that quantified the quality of the energy
reconstruction on an event-by-event basis, since it is expected to improve the
statistical power of the search (see section~IV of
ref.~\cite{Fermi-LAT:2013uma}).
We found that while this improvement was important in searches for
higher-energy spectral lines ($\sim 15\%$ increase in statistical power for
$E\gtrsim 10$\,GeV), it was less than 10\% at lower energies, and is hence
neglected in the present analysis.  At higher energies
the quality of the energy reconstruction can vary markedly from event to
event.  For example, an on-axis 100\,GeV \gammaRayHyph\ event
leaks about 50\% of its energy out the back of the \Fermi\/-LAT~\cite{REF:2012.P7Perf}.
Events at this energy with larger incident angles will travel
through more radiation lengths in the \Fermi\/-LAT calorimeter, leak less
energy out the back, and therefore typically have
more accurate energy reconstructions.  However, the difference between the
quality of the energy reconstruction is less dramatic at lower energies
($\lesssim 1$\,GeV), where the energy deposition is usually fully contained
in the \Fermi\/-LAT calorimeter.
\medskip

The most relevant effect of any systematic biases that masks or fakes a line-like signal
(this includes both instrumental effects as well as effects due to the power-law
approximation of the background spectra) is to offset the estimated number of signal
events with respect to its true value.  We expect such offsets to
scale linearly with the number of events in the ROI; therefore it is useful to introduce
the \textit{fractional deviation} $f$, which, roughly speaking, denotes the
fractional size of a line signal relative to the background under the
signal peak (similar to signal-to-background ratio):
\begin{linenomath}
\begin{equation}
    f\equiv n_{\text{sig}} / b_{\text{eff}}\,,
    \label{equ:fracDev}
\end{equation}
\end{linenomath}
where $b_{\text{eff}}$ denotes the number of effective background events below a
line signal. 
For each ROI and value of
$E_\gamma$, the number of effective background events is obtained as
\begin{linenomath}
\begin{equation}
    b_{\text{eff}} = \int_{E_i^-}^{E_i^+} dE\, \frac{\mathcal{D}_{\text{eff}}(E|E_\gamma)\alpha
        E^{-\Gamma}\mathcal{E}(E)}{\alpha
        E^{-\Gamma}\mathcal{E}(E) +
    \mathcal{D}_{\text{eff}}(E|E_\gamma)}\,,
    \label{eqn:beff}
\end{equation}
\end{linenomath}
where $\alpha$ and $\gamma$ are determined from a power-law only fit to the data (with
$n'_{\text{sig}}=0$ fixed).

A systematic uncertainty in the number of signal events can now be conveniently
expressed as being proportional to the fractional deviation, $\delta f$.
For most systematically induced features that could fake or hide a line signal, this quantity will
be approximately independent of the number of measured events in the adopted
ROI.  The corresponding distribution function, $P_\mathcal{F}(n_{\text{syst}},
b_\text{eff})$ in
eq.~\eqref{eqn:likelihood}, will be determined empirically as discussed in the
next subsection.

\medskip

As usual, upper limits at the $95\%$ confidence level (CL) on the number of signal events $n_{\text{sig}}$ 
are obtained by increasing $n_{\text{sig}}$, while refitting all other
parameters, until $-2\ln \mathcal{L}$ changes by $2.71$ from its best-fit
value.  The significance of a line signal in units of Gaussian sigma is given
by $\sqrt{2\ln\mathcal{L}/\mathcal{L}_0}$, where
$\mathcal{L}_0$ denotes the likelihood of a fit with the line flux set to zero.  
Note that in our analysis, we neglect corrections to the finite angular
resolution of the \Fermi\/-LAT.  Furthermore, by construction, effects related
to modelling uncertainties (i.e.\ modelling the effective area, background
emission, and not masking known point sources) are absorbed in $P_{\mathcal{F}}(n_{\text{syst}}, b_\text{eff})$.  

\subsection{Systematics}\label{sec:syst}
As discussed in section~VI of ref.~\cite{Fermi-LAT:2013uma}, there are three classes
of systematic uncertainties involved when searching for \gammaRayHyph\ spectral
lines: uncertainties on the calculated exposure ($\delta\mathcal{E} /
\mathcal{E}$), uncertainties on the fit estimates of the signal counts ($\delta
n_{\text{sig}} / n_{\text{sig}}$), and line-like uncertainties that could mask a true signal
or  induce a false signal ($\delta f$) that we discussed in the previous subsection. 
The two former are less worrisome since they are smaller than the statistical
uncertainty on the 95\% CL limit on $n_{\text{sig}}$ ($\sim 50\%$ since $n_{\text{sig}} \ll b_\text{eff}$, causing the statistical uncertainty on $n_{\text{sig}}$ to be $\simeq\sqrt{b_\text{eff}}$), and
can safely be neglected.

The latter systematic uncertainties are especially worrisome since
positive features could induce false signals, while negative features could
mask true signals. We quantify these in terms of an uncertainty on the
fractional deviation (see eq.~(\ref{equ:fracDev})), $\delta f$. The statistical uncertainty is $\delta f_{\text{stat}} \simeq 1/\sqrt{b_{\text{eff}}}$, while systematically induced fractional deviations are expected to be $\delta f_{\text{syst}}\simeq{\text{constant}}$.  Therefore, as $b_{\text{eff}}$ increases (i.e.\ the number of events used in the fit increases), the systematic uncertainties can begin to dominate ($\delta f_{\text{stat}} \ll \delta f_{\text{syst}}$).  This is the case for all of our low energy fits ($E_{\gamma}\lesssim3$\,GeV), which is why it is
necessary to include systematic uncertainties correctly in the fitting
procedure.

As mentioned in the previous subsection, we incorporate the systematic
uncertainties into our likelihood formalism via $P_\mathcal{F}(n_{\text{syst}},
b_\text{eff})$ (see eq.~\eqref{eqn:likelihood}).  We break the degeneracy between $n_{\text{syst}}$ and $n_{\text{sig}}$ by  constraining $n_{\text{syst}}$ with a Gaussian distribution\footnote{We also studied modeling $n_{\text{syst}}$ with top hat and triangle functions with a base width of $2\delta f_{\text{stat}}$.  They improved and worsened the limits by $\sim30\%$ respectively.  Given our choice of $\delta f_{\text{stat}}=0.011$, we consider this modeling choice to be simple, but conservative.}
\begin{linenomath}
\begin{equation}
    P_\mathcal{F}(n_{\text{syst}}, b_\text{eff}) =
    \frac{1}{\sigma_{\text{syst}}\sqrt{2\pi}}\exp\left(-\frac{(n_{\text{syst}}-\mu_{\text{syst}})^2}{2\sigma_{\text{syst}}^{2}}\right).
    \label{eqn:Pf_sys}
\end{equation}
\end{linenomath}
We chose to set $\mu_{\text{syst}}=0$ and define $\sigma_{\text{syst}}=\delta
f_{\text{syst}}b_{\text{eff}}$, where $\delta f_{\text{syst}}$ was determined based on fits
for line-like signals in control regions. One could model $n_{\text{syst}}$
more aggressively, for example in an energy-dependent way, but we chose not to
since we have only a limited number of control regions available to
verify the energy dependence of $n_{\text{syst}}$.

We fit for line-like signals in control regions where we do not expect any DM signal to dominate in
order to estimate $\delta f_{\text{syst}}$. We scan in 0.25$\sigma_E$ steps in energy for line-like signals (allowing for both positive and negative
signals) in $20^{\circ}\times20^{\circ}$ ROIs along the Galactic plane in
$10^{\circ}$ steps excluding the 5 centre-most ROIs (i.e.\ $|b|>20^{\circ}$; 31
total ROIs; cf.\ figure~\ref{fig:rois}). Since the DM signal
is expected to peak in the Galactic Centre, this is a control region where non-DM
astrophysical processes dominate the observed \gammaRayHyph\ emission. Systematically induced line-like features will result from modelling imperfections like averaging the
energy-dependent variations in the \Fermi\/-LAT effective area over the ROI,
not masking or modelling known point sources, and modelling the background
spectrum as a power law.  It is not possible to disentangle these components in our Galactic plane scans, so we consider them together as modelling imperfections. We also studied the fractional deviations observed in \gammaRays\ from the Earth's limb emission and the Vela pulsar, see appendix~\ref{app:control_samples}.

Figure~\ref{fig:GalRidge} shows the fractional deviations observed in the Galactic
plane scan. Also shown is the average statistical uncertainty of the fractional
deviation.  If there were no systematic effects, one
would expect $\delta f_{\text{stat}}$ to contain 68\% of the observed
fractional deviations.  Clearly this is not the case, especially at lower
energies, showing that systematic effects are not negligible.  At high
energies, $\gtrsim3$\,GeV, the fits are dominated by statistical variations,
while at lower energies the fits are dominated by systematic effects.
We calculated the $\delta f$ values that contained 68\% of the Galactic plane
fits, $\delta f_{68}(E)$, in a small energy range ($\pm 10\%$).  To be
conservative, we choose the largest $\delta f_{68}$ value observed in the
Galactic plane scan (for $E_{\gamma}<3$\,GeV) as our estimate for the
systematic uncertainty from 
biases in our modelling of the LAT effective area, point-source contributions,
and the background spectral shape; $\delta f_{\text{GP}}=0.011$. 

From figure~\ref{fig:GalRidge}, we can infer some properties of the systematic
uncertainties that affect our search. The displacement of $\delta f$ from zero and common variations with energy between all the control ROIs are most likely caused by small biases in modelling the \Fermi\/-LAT effective area.  The spread amongst the fits in the control
ROIs is probably from our modelling of the background spectra by a power law.

\begin{figure}[t]
  \centering
  \includegraphics[width=0.6\twocolfigwidth]{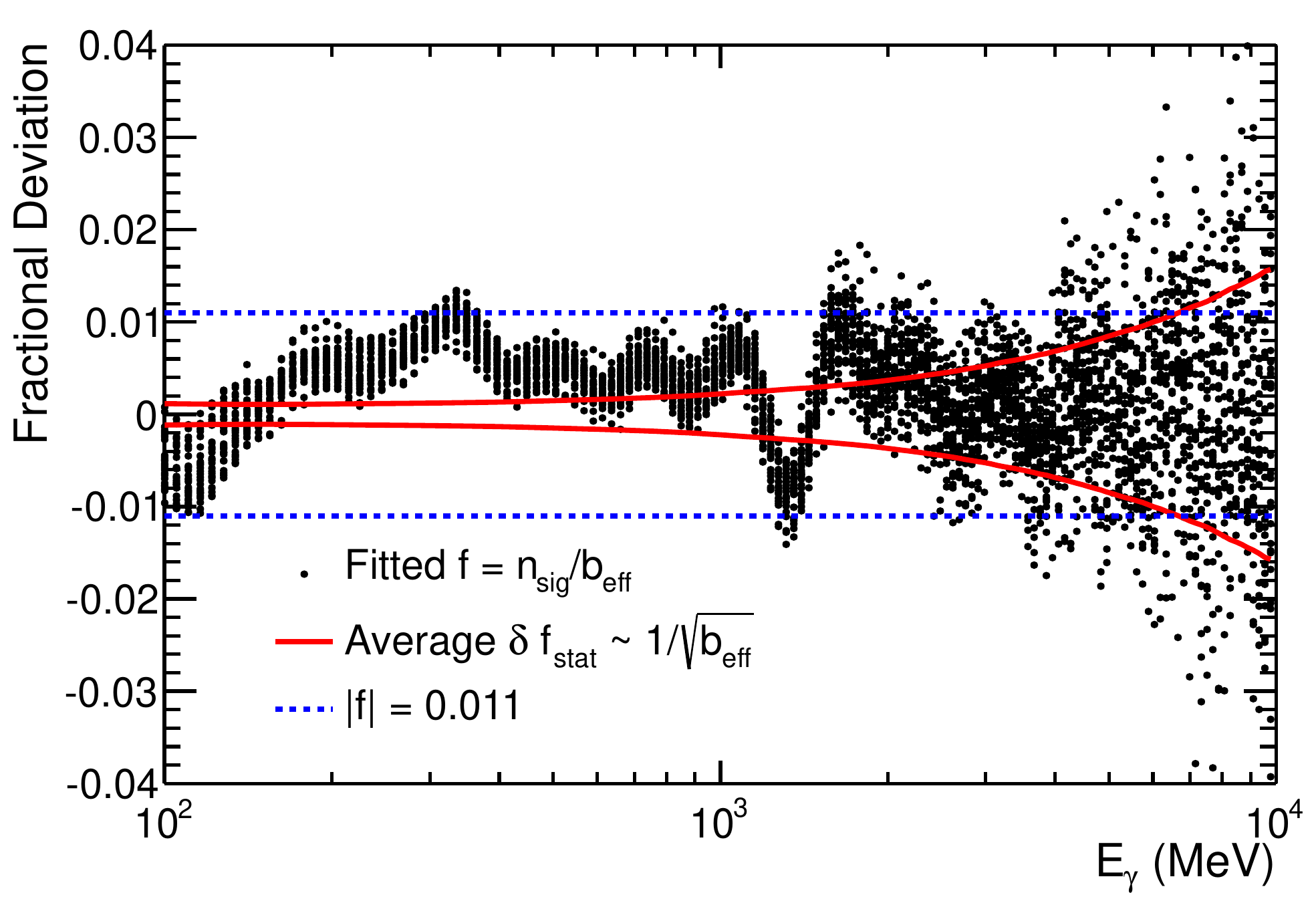}
  \caption{\label{fig:GalRidge}Fractional deviations ($f$, see eq.~(\ref{equ:fracDev})) observed in the Galactic plane scan are shown as black dots.  $E_{\gamma}$ went from 100\,MeV to 10\,GeV in steps of $0.25\sigma_E$. The red line shows the average statistical uncertainty from the Galactic plane scan.  The blue dashed line shows the value we chose to represent the $\delta f$ from modelling biases; see text for details.}
\end{figure}

We also estimate the systematic uncertainty from residual cosmic-ray events passing our \gammaRayHyph\ event selection. Since we use the \irf{P7REP\_CLEAN}\ event class, the cosmic-ray contamination is not expected to be a large effect, especially for the region ROI$_{\text{cen}}$, which focuses on the bright Galactic Centre.  However, cosmic-ray contamination is worrisome at large latitudes (e.g.\ ROI$_{\text{pol}}$ region). To study the effect of cosmic-ray contamination, we select events that are included in the less stringent \irf{P7REP\_SOURCE}\ class, but are not included in the \irf{P7REP\_CLEAN}\ class in the ROI$_{\text{pol}}$ region.  This sample will be enriched with cosmic-ray events that were not removed by the \irf{P7REP\_SOURCE}\ selection, but did not pass the \irf{P7REP\_CLEAN}\ event selection. Similar to what was done in ref.~\cite{Fermi-LAT:2013uma}, we take the largest observed $\delta f$ in this control sample along with the expected \gammaRayHyph\ acceptance ratio between the \irf{P7REP\_CLEAN}\ and \irf{P7REP\_SOURCE}\ selections (see appendix~D5 in ref.~\cite{Fermi-LAT:2013uma}) to obtain an estimate of $\delta f_{\text{CR}}\sim0.001$ and $\delta f_{\text{CR}}\sim0.01$ in ROI$_{\text{cen}}$ and ROI$_{\text{pol}}$ respectively. Therefore we obtain a final estimate of $\delta f_{\text{syst}}=0.011$ and $\delta f_{\text{syst}}=0.015$ in ROI$_{\text{cen}}$ and ROI$_{\text{pol}}$ respectively.

Other systematic uncertainties in this search enter from our calculation of the
\Fermi\/-LAT exposure, modelling of the energy dispersion, and our choice of
$E_{\gamma}$ grid spacing.  The overall uncertainty in the calculation of the
\Fermi\/-LAT effective area is $\sim10\%$.  Additionally, we choose to use the
average exposure across our ROIs when converting from counts to flux. The
\Fermi\/-LAT observes the sky with relative uniform exposure, but it does vary
by $\delta\mathcal{E} / \mathcal{E} = 0.02$ in ROI$_{\text{cen}}$ and by $\delta\mathcal{E} / \mathcal{E} = 0.07$ in ROI$_{\text{pol}}$.  When added in
quadrature, we have a total systematic uncertainty on the exposure of
$\delta\mathcal{E}/\mathcal{E} = 0.10$ and $\delta\mathcal{E}/\mathcal{E} = 0.12$ in ROI$_{\text{cen}}$ and ROI$_{\text{pol}}$ respectively.  Additionally,  we estimate the effect of the $10\%$ uncertainty in the energy resolution~\cite{REF:2012.P7Perf} to be $\delta n_{\text{sig}} / n_{\text{sig}} \simeq 7\%$.  Also, fitting at fixed $E_{\gamma}$ values with a grid spacing of $0.5\sigma_E$ would cause us
to undermeasure the number of signal counts by 10\% at most if the true line
were in between two fit points. Therefore, the total uncertainty on the number
of signal counts is $\delta n_{\text{sig}}/n_{\text{sig}}=^{+0.07}_{-0.12}$. These systematic
uncertainties, however, would simply scale our limits up or down by the
reported uncertainties, which are smaller than the expected statistical variation of the limits ($\sim50\%$), 
and hence neglected.

\subsection{Results}\label{sec:Results}
We scanned for \gammaRayHyph\ lines from
$100\,\text{MeV}$ to $10$\,GeV in both ROI$_{\text{cen}}$ (annihilation-optimised)
and ROI$_{\text{pol}}$ (decay-optimised) and find no significant
detections.  Note that all our fits had signals with a significance less than $1\sigma$, which is likely an indication that our assignment of $\delta f_{\text{syst}}$ is rather conservative.  As discused above, independent verification of both the magnitude and energy dependence of $\delta f_{\text{syst}}$ is not available.  Therefore we chose to simply treat $\delta f_{\text{syst}}$ as a constant determined by fits in our control regions (see section~\ref{sec:syst}).

We set 95\% CL upper limits on $n_{\text{sig}}$ using
the method described at the end of section~\ref{sec:FitProc} at each
energy in both ROIs. For a monochromatic signal we can convert the
$n_{\text{sig}}$ upper limits to flux upper limits using the ROI-averaged
exposure $\mathcal{E}_{\text{ROI}}(E_{\gamma})$:
\begin{linenomath}
\begin{equation}
    \Phi_{\text{mono}}(E_{\gamma})=\frac{n_{\text{sig}}(E_{\gamma})}{\mathcal{E}_{\text{ROI}}(E_{\gamma})}\,. 
    \label{equ:FluxFromCounts}
\end{equation}
\end{linenomath}
Figure~\ref{fig:FluxLim} shows the 95\% CL $\Phi_{\text{mono}}$ upper limits obtained in
ROI$_{\text{pol}}$ and ROI$_{\text{cen}}$.
We show both the limits obtained assuming no systematic uncertainties
($\delta f_{\text{syst}}=0$) and those obtained including the appropriate $\delta f_{\text{syst}}$ determined
by fits in our control regions. 

% created these plots with Limits_Flux.py
\twopanel{t}{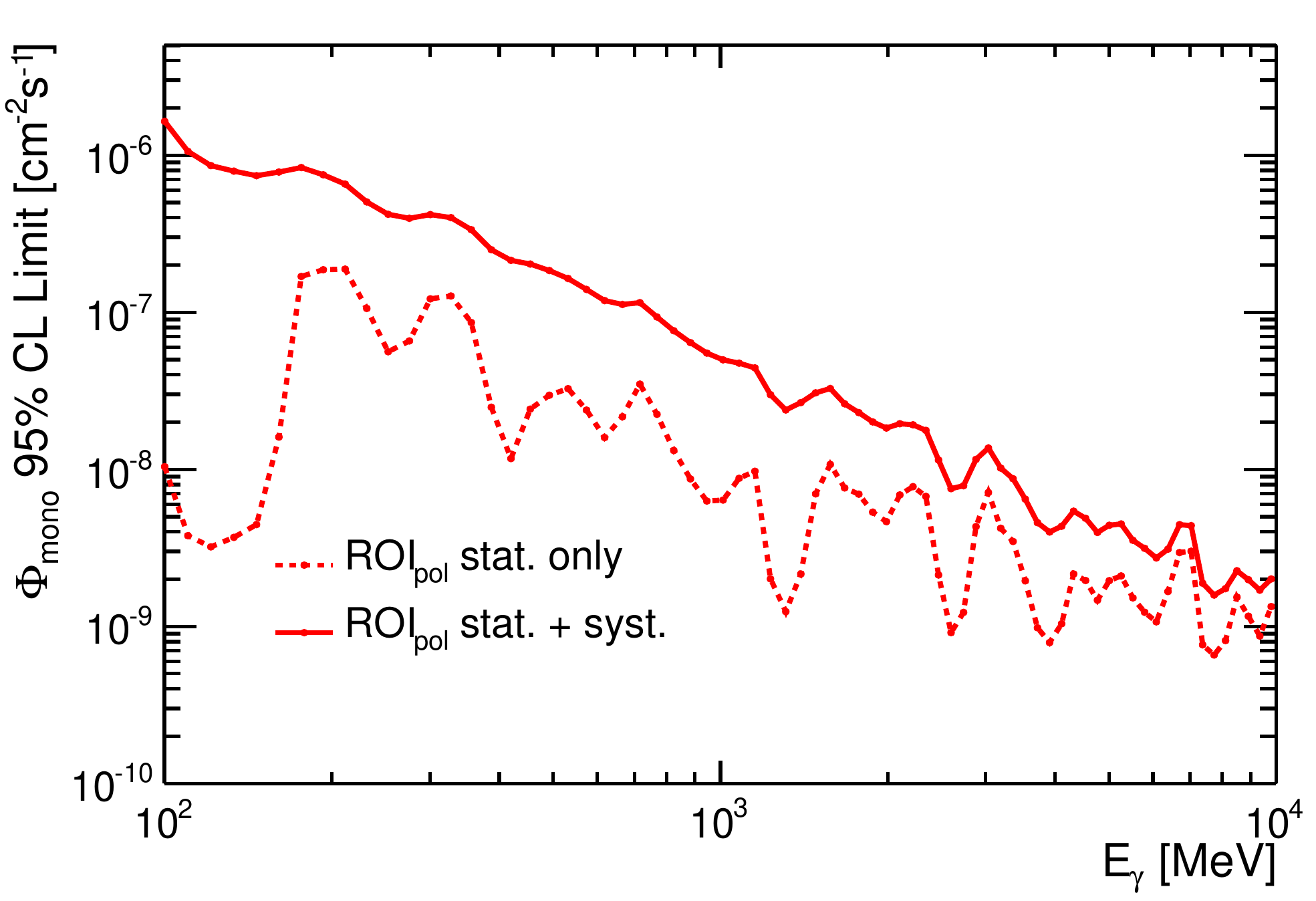}{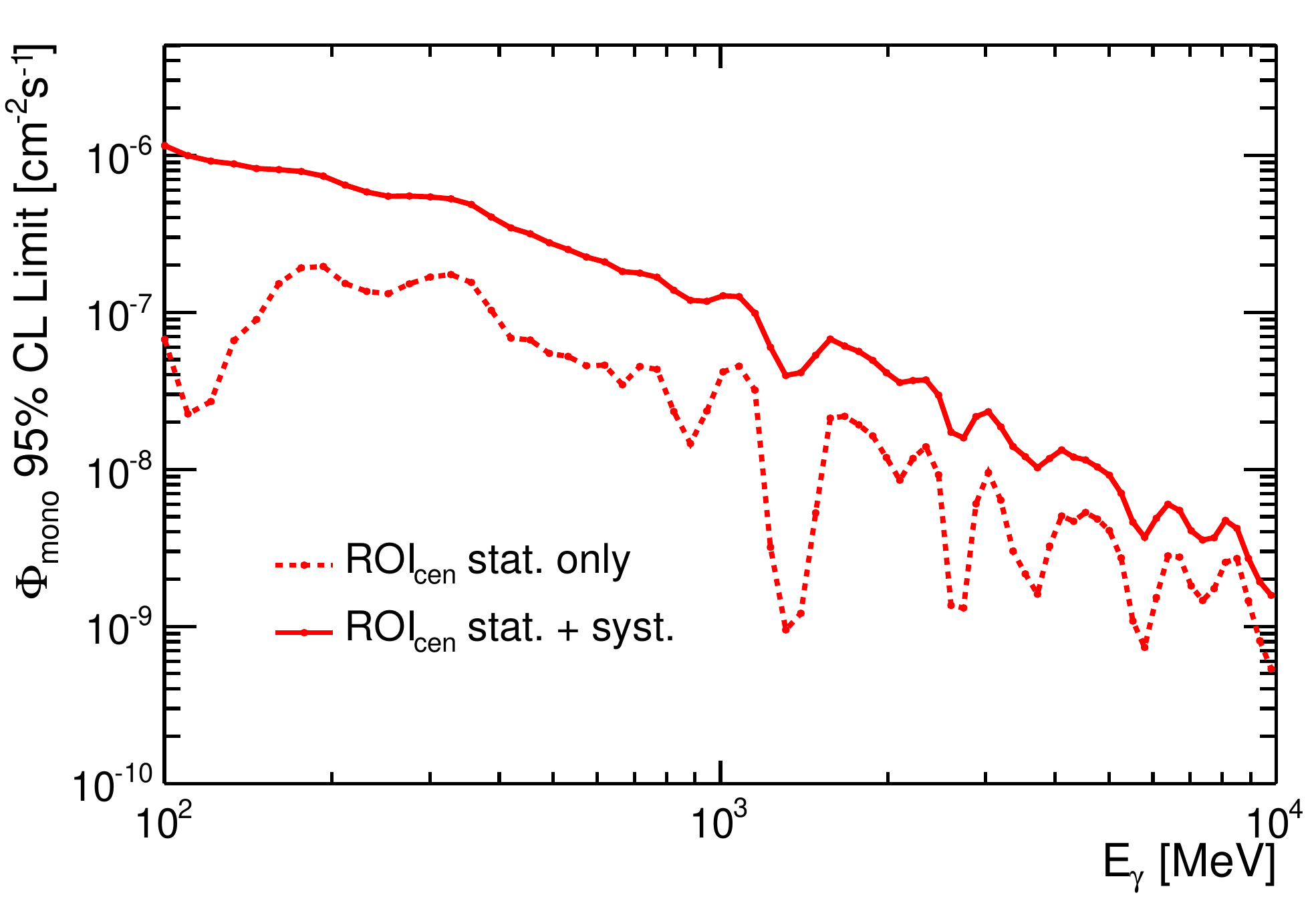}
{\caption{\label{fig:FluxLim}95\% CL $\Phi_{\text{mono}}$ upper limits in ROI$_{\text{pol}}$
(left) and ROI$_{\text{cen}}$ (right). The solid line shows the limits obtained using $\delta
f_{\text{syst}}$ values determined from fits in our control regions.  The
dashed line shows the limits obtained neglecting the systematic uncertainties
($\delta f_{\text{syst}}=0$).}}

Assuming the monochromatic signal is coming completely from either DM decay or
annihilation, we can find the 95\% CL lower limits for $\tau_{\gamma\nu}$ (see
eq.~(\ref{eq:decayFlux})) and upper limits for $\langle\sigma
v\rangle_{\gamma\gamma}$ (see eq.~(\ref{eq:annihilationFlux})), respectively.
These limits are shown in figure~\ref{fig:DMLim}.  Furthermore, we provide in
appendix~\ref{app:tables} tables of the flux upper limits, the lifetime lower
limits and the annihilation cross section upper limits. 

% created these plots with Limits_DM.py
\twopanel{t}{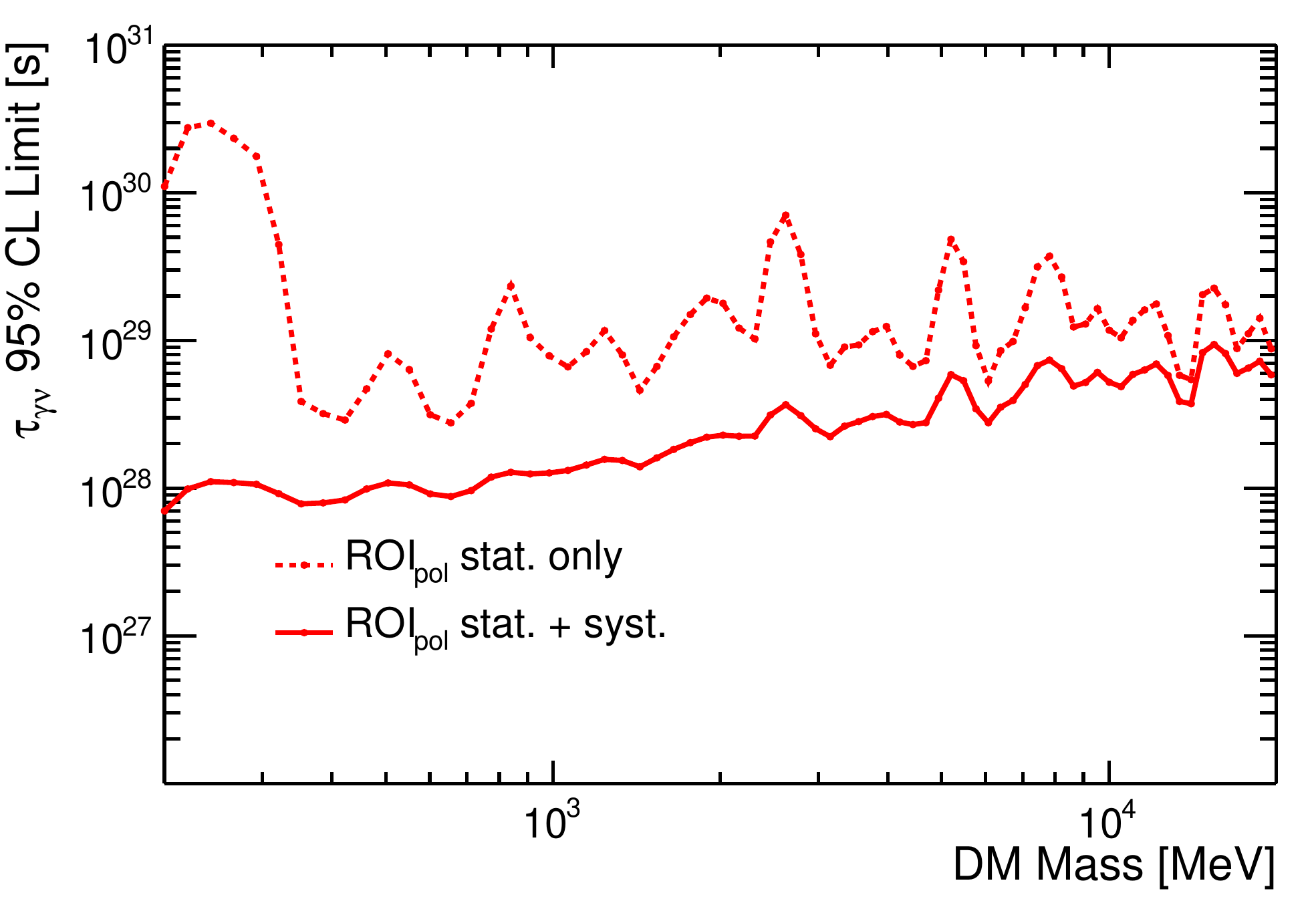}{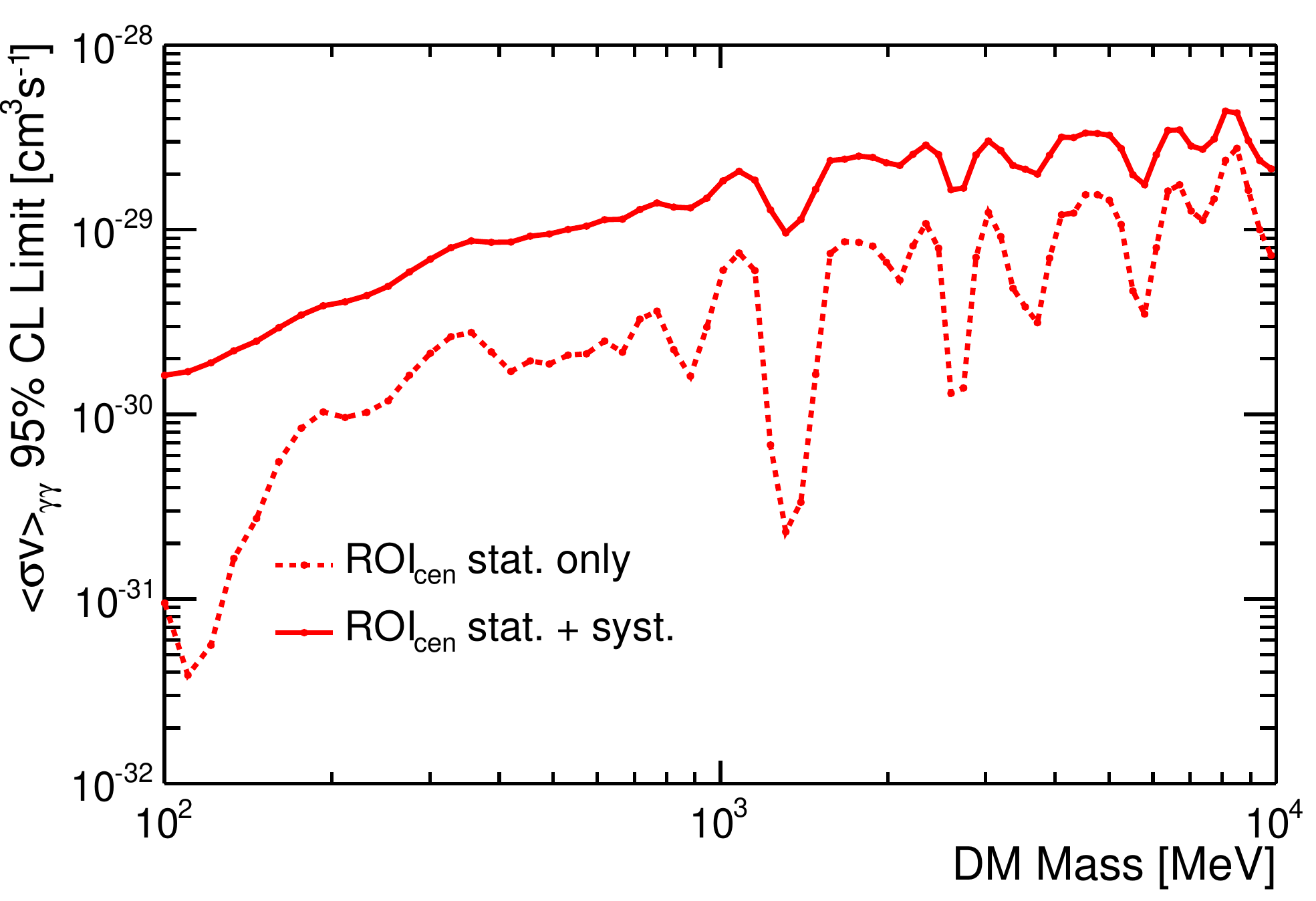}
{\caption{\label{fig:DMLim}95\% CL $\text{DM}\rightarrow\gamma\nu$  decay
lifetime ($\tau_{\gamma\nu}$) lower limits (left) and $\text{DM DM}\to
\gamma\gamma$ cross section ($\langle\sigma v\rangle_{\gamma\gamma}$) upper
limits (right). The solid line shows the limits obtained using $\delta f_{\text{syst}}$ values determined from fits in our control regions.  The dashed line shows the limits obtained neglecting the systematic uncertainties ($\delta f_{\text{syst}}=0$).}}

\section{Discussion}
\label{sec:discussions}
\subsection{\texorpdfstring{Implications for \boldmath$\mu\nu$SSM gravitino dark matter}{Implications for munuSSM gravitino dark matter}}\label{results}

\begin{figure}[t]
  \centering
  \includegraphics[width=0.8\twocolfigwidth]{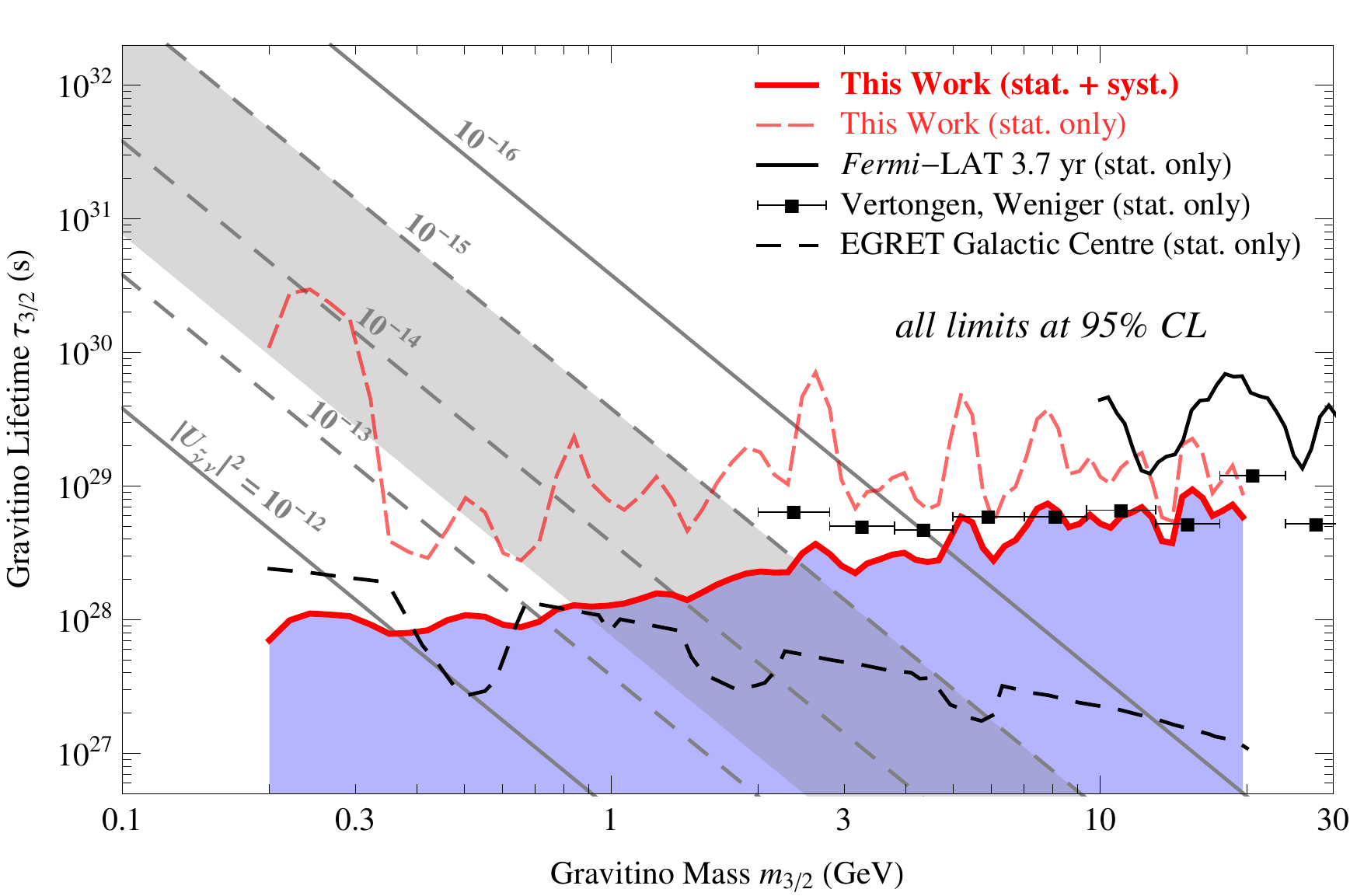}
  \caption{Parameter space of decaying gravitino DM given in terms of the gravitino lifetime and the gravitino mass. The diagonal band shows the allowed parameter space for gravitino DM in the $\mu\nu$SSM. The numbers on the solid and dashed lines show the corresponding value of the photino--neutrino mixing parameter, as discussed in section~\ref{munussm}. The theoretically most favoured region is coloured in grey. We also show several 95\% CL lower limits on the gravitino lifetime coming from \gammaRayHyph\  observations. The blue shaded region is excluded by the limits derived in this work.}
  \label{fig:lifetime}
\end{figure}

\begin{figure}[t]
  \centering
  \includegraphics[width=0.8\twocolfigwidth]{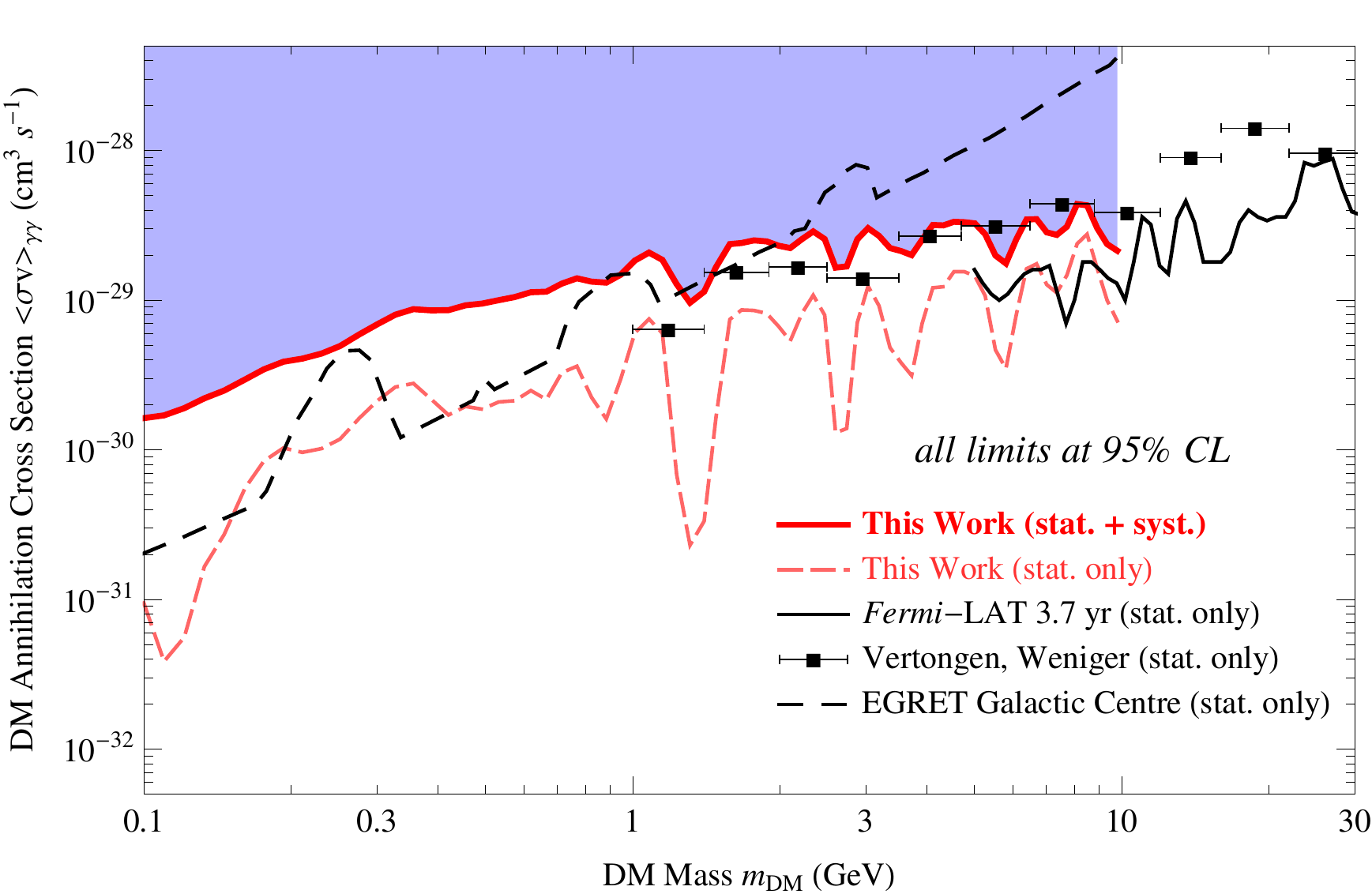}
  \caption{Comparison of the 95\% CL upper limits on the DM pair annihilation cross section into two photons found in this work to earlier results using \Fermi\/-LAT and EGRET data. The blue shaded region is excluded by the limits derived in this work.}
  \label{fig:crosssection}
\end{figure}

Let us finally apply the results obtained above to the case of gravitino DM in
the framework of the $\mu\nu$SSM, assuming that the gravitino constitutes 100\% of the DM in the Universe. 
In figure~\ref{fig:lifetime} we show the
parameter space for decaying gravitino DM in terms of $\tau_{3/2}$ and
$m_{3/2}$ (see figure~\ref{fig:crosssection} for the corresponding plot for
annihilation signals). The $\mu\nu$SSM prediction for the parameter range is shown as a diagonal band bounded by solid lines (see eq.~(\ref{munuSSMlifetime})).
As discussed in section~\ref{munussm}, the photino--neutrino mixing parameter is constrained to be in the range shown in eq.~(\ref{representative}) in order to reproduce the neutrino masses correctly. 
As a consequence, any acceptable set of gravitino parameters must lie within the diagonal solid lines.
The favoured range for the photino--neutrino mixing parameter, eq.~(\ref{scan}), is coloured in grey in the figure. 

The stat.\ + syst.\ limit (red thick solid line of figure~\ref{fig:lifetime}) excludes at 95\% CL values of $m_{3/2}$ in the $\mu\nu$SSM larger than 4.8\,GeV and restricts
$\tau_{3/2}$ to be larger than at least $7.9\times 10^{27}$\,s for lower gravitino
masses within the mass range probed by our analysis. Considering the favoured
range (grey band), this 95\% CL limit implies $m_{3/2}$ to be below 2.4\,GeV and $\tau_{3/2}$ to be 
larger than at least $1.3\times 10^{28}$\,s for lower gravitino
masses. It is worth noting that the stat.\ + syst.\ limit is the most robust current bound since we are
considering the most relevant systematic effects that may enhance and/or fake a
gravitino decay signal.  Furthermore, the uncertainty in the DM distribution within the
ROI$_{\text{pol}}$ target region is rather small (less than $\sim 10\%$) within the context of the local DM density and various DM
profiles we consider. 

The above results also allow us to discard at 95\% CL a large fraction of the $\mu\nu$SSM parameter space
($m_{3/2}$, $\tau_{3/2}$) presented in~\cite{gustavo}, where
a gravitino signal was predicted to be detectable through observations of the Virgo galaxy cluster after 5 years of \Fermi\/-LAT operation.

% created this plot with MathematicaPlots.nb

% created this plot with MathematicaPlots.nb

\subsection{Comparison with previous limits and results}

Limits on \gammaRayHyph\ line emission in the energy range 100\,MeV to 10\,GeV were in
the past derived by a number of groups, both for the case of DM decay
and for DM annihilation.  For comparison, the limits from EGRET observations of the Galactic
Centre~\cite{EGRET}\footnote{We adopted the limits on the \gammaRayHyph\ flux from the dashed line in figure~7 of ref.~\cite{EGRET} (sliding window technique) and calculated the constraints on the lifetime and the annihilation cross section for the case of the Einasto profile adopted in this work. For the $10^\circ\times10^\circ$ ROI around the Galactic Centre with size $\Delta\Omega_{\text{ROI}}=0.0304\,$sr used in the EGRET analysis, the $J$-factors are given by $0.522\times 10^{22}\,\text{GeV\,cm}^{-2}$ for decay and $5.33\times 10^{22}\,\text{GeV}^2\,\text{cm}^{-5}$ for annihilation, respectively.}, limits using \Fermi\/-LAT measurements in a number of ROIs separately optimised for DM decay and DM annihilation~\cite{Fermi-LAT:2013uma},\footnote{The limits are taken from tables~VIII to~X of ref.~\cite{Fermi-LAT:2013uma}. The cross section limits are those from the ROI R16 optimised for the Einasto profile and the lifetime limits are those from the ROI R180 rescaled by a factor $2.49/2.46=1.01$ to account for the $J$-factor of the Einasto profile as given in table~II of ref.~\cite{Fermi-LAT:2013uma}.} and limits derived from \Fermi\/-LAT data in individual energy ranges~\cite{Vertongen:2011mu}\footnote{The limits were taken from table~3 of ref.~\cite{Fermi-LAT:2013uma} and rescaled with the appropriate factor for the Einasto profile as given in table~2 of ref.~\cite{Fermi-LAT:2013uma}.} are also shown in figures~\ref{fig:lifetime} and~\ref{fig:crosssection}. Care has to
be taken when comparing these older results with ours, since our limits take
systematic effects into account for the first time, whereas all previous
analyses were based on statistical errors only.  As a consequence, our final
limits are --- where they overlap --- a factor 2--11 (decay) or 2--4 (annihilation) weaker than the limits
derived in the high-energy line analysis in ref.~\cite{Fermi-LAT:2013uma}, and up to a factor
of 3 weaker (decay and annihilation) than the conservative limits from ref.~\cite{Vertongen:2011mu}.  At energies
below roughly 400\,MeV (decay) or below roughly 1\,GeV (annihilation), our limits are furthermore slightly weaker than previous results
from EGRET.\footnote{The ROI used in the EGRET analysis was considerably smaller than ROI$_\text{cen}$. However, recall that we chose to use a $20^\circ \times 20^\circ$ ROI in order to avoid effects from the LAT PSF though the optimal ROI was smaller.}  For comparison we show in figures~\ref{fig:lifetime} and~\ref{fig:crosssection} the results that we obtain when
setting the fractional deviation systematics to zero.  In that case, our limits
become --- as expected for the large effective area of the \Fermi\/-LAT --- nominally
stronger by up to an order of magnitude, and they are in particular stronger
than most previous results.
\section{Conclusions}\label{conclusions}

In this work we searched for \gammaRayHyph\ lines from 100\,MeV to 10\,GeV using 5.2 years of \Fermi\/-LAT data.  We expect these low-energy spectral lines from decaying gravitino DM, but also extend our search to including DM annihilations into a pair of \gammaRays.  We did not find any statistically significant spectral lines and have set robust limits on DM interactions that would produce monochromatic \gammaRays.

Given the large number of events in our fits, most were dominated by systematic uncertainties since the statistical uncertainties became very small.  Therefore, it was critical to appropriately include the systematic uncertainties in our likelihood formalism.  For the first time, we present robust limits for monochromatic \gammaRays\ that incorporate systematic uncertainties.  We conservatively determine the level of systematic uncertainties from fits to control regions where no line-like signals are expected.  While our limits are not much
more constraining than previous limits, they are more robust.

We discussed the results in the context of the $\mu\nu$SSM, a Supersymmetric Standard Model that simultaneously solves the $\mu$ problem and explains neutrino masses and mixing angles by the addition of right-handed neutrinos to the theory. The gravitino is a well-motivated candidate for the DM as it can have the correct relic density (cf.\ figure~\ref{fig:relicdensity}) and leads to a consistent cosmological scenario.
As a consequence of the \gammaRayHyph\ line search results, the gravitino DM mass in the $\mu\nu$SSM must be $m_{3/2}<4.8$\,GeV and the lifetime $\tau_{3/2}>7.9\times10^{27}$\,s, at 95\% CL if we assume that all of the DM in the Universe is in the form of gravitinos.  In the favoured model parameter space these limits tighten to $m_{3/2}<2.4$\,GeV and $\tau_{3/2}>1.3\times10^{28}$\,s (see figure~\ref{fig:lifetime}).

\acknowledgments

The \Fermi\/-LAT Collaboration acknowledges generous ongoing support
from a number of agencies and institutes that have supported both the
development and the operation of the \Fermi\/-LAT as well as scientific data analysis.
These include the National Aeronautics and Space Administration and the
Department of Energy in the United States, the Commissariat \`a l'Energie Atomique et aux Energies Alternatives
and the Centre National de la Recherche Scientifique / Institut National de Physique
Nucl\'eaire et de Physique des Particules in France, the Agenzia Spaziale Italiana
and the Istituto Nazionale di Fisica Nucleare in Italy, the Ministry of Education,
Culture, Sports, Science and Technology (MEXT), High Energy Accelerator Research
Organization (KEK) and Japan Aerospace Exploration Agency (JAXA) in Japan, and
the K.~A.~Wallenberg Foundation, the Swedish Research Council and the
Swedish National Space Board in Sweden.

Additional support for science analysis during the operations phase is gratefully
acknowledged from the Istituto Nazionale di Astrofisica in Italy and the Centre National d'\'Etudes Spatiales in France.\smallskip

We thank Luca Baldini, Philippe Bruel, Seth Digel, Michael Gustafsson, Daniel E.\ L\'opez-Fogliani, and Miguel \'A.\ S\'anchez-Conde for useful discussions and comments.\smallskip

The work of GAGV was supported by Conicyt Anillo grant ACT1102. GAGV and CM thank for the support of the Spanish MINECO's Consolider-Ingenio 2010 Programme under grant MultiDark CSD2009-00064. Their work was also supported in part by MINECO under grant FPA2012-34694. GAGV, MG and CM acknowledge the support of the Marie Curie ITN ``UNILHC'' under grant number PITN-GA-2009-237920, the support by the Comunidad de Madrid under grant HEPHACOS S2009/ESP-1473, and the support of the MINECO under the ``Centro de Excelencia Severo Ochoa'' Programme SEV-2012-0249. MG also thanks for the support of the Forschungs- und Wissenschaftsstiftung Hamburg through the program ``Astroparticle Physics with Multiple Messengers'' and the Marie Curie ITN ``INVISIBLES'' under grant number PITN-GA-2011-289442.

\clearpage

% Specify following sections are appendices. 
\appendix

\section{Choice of the dark matter halo profile}\label{app:halo_profile}

\begin{figure}[t]
\centering
  \includegraphics[width=0.47\linewidth]{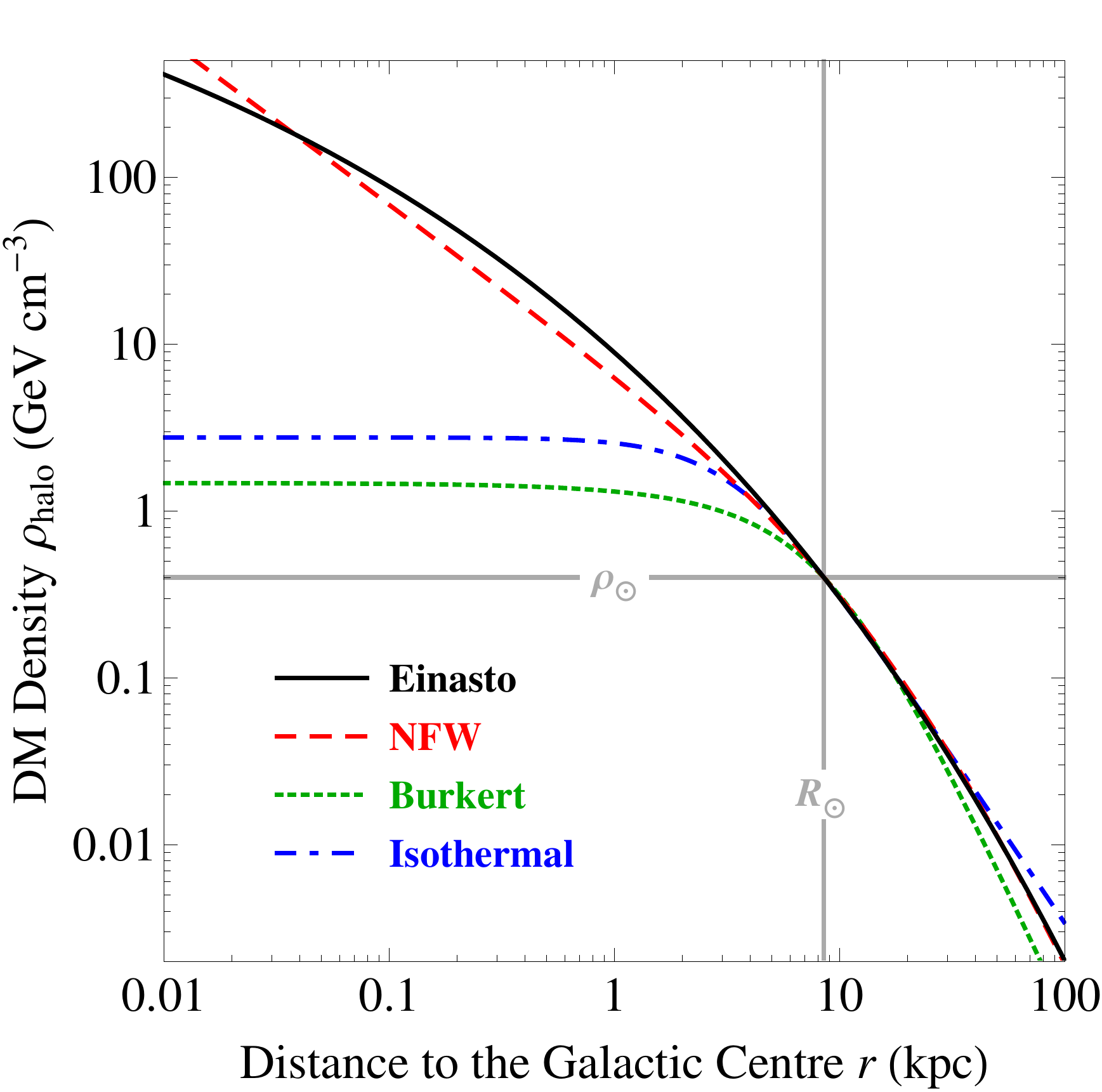}
  \hfill
  \caption{Comparison of different DM density profiles for the Galactic halo.}
  \label{fig:profiles}
\end{figure}

The distribution of the DM in the Milky Way is not known and therefore presents a source of uncertainty to any analysis of \gammaRayHyph\ signals from DM decay or annihilation. N-body cosmological simulations favour cuspy profiles like the Einasto profile introduced in the main text or the NFW profile~\cite{Navarro:1995iw}:
\begin{linenomath}
\begin{equation}
 \rho_{\text{NFW}}(r)=\rho_{\odot}\,\frac{\left( R_{\odot}/r_s\right) \left( 1+R_{\odot}/r_s\right) ^2}{\left( r/r_s\right) \left( 1+r/r_s\right) ^2}\,,
\end{equation}
\end{linenomath}
where $r_s\simeq 20\,$kpc for the case of the Milky Way. By contrast some observations at other halo mass scales favour cored profiles like the isothermal profile~\cite{Bahcall:1980fb}:
\begin{linenomath}
\begin{equation}
 \rho_{\text{iso}}(r)=\rho_{\odot}\,\frac{1+\left( R_{\odot}/r_s\right) ^2}{1+\left( r/r_s\right) ^2}\,,
\end{equation}
\end{linenomath}
where $r_s\simeq 3.5\,$kpc for the case of the Milky Way halo~\cite{Bertone:2004pz}, or the Burkert profile~\cite{Burkert:1995yz}:
\begin{linenomath}
\begin{equation}
 \rho_{\text{Bur}}(r)=\rho_{\odot}\,\frac{\left(1+\left( R_{\odot}/r_s\right) \right)\left(1+\left( R_{\odot}/r_s\right) ^2\right)}{\left(1+\left( r/r_s\right) \right)\left(1+\left( r/r_s\right) ^2\right)}\,,
\end{equation}
\end{linenomath}
where $r_s\simeq 9\,$kpc~\cite{Nesti:2013uwa}. In figure~\ref{fig:profiles} we compare 
the different DM density profiles. In all cases we fixed the normalisation to a local DM density of
$\rho_{\odot}\simeq0.4\,\text{GeV}\,\text{cm}^{-3}$~\cite{Catena:2009mf,Weber:2009pt,Salucci:2010qr}, 
keeping in mind that this value is rather uncertain and the true value could be up to a factor of 2 lower or higher 
(see for instance ref.~\cite{Cirelli:2010xx} and references therein). 
Moreover, the value that best fits observational data may depend on the choice of the DM density profile~\cite{Iocco:2011jz,Gomez-Vargas:2013bea}. 
While all these profiles behave similarly in the outer part of the Milky Way, they
deviate significantly in the vicinity of the Galactic Centre.
%
% created this plot with MathematicaPlots.nb
The $J$-factors that we obtain for the different profiles in our ROIs are
summarised in table~\ref{tab:jfactors}.  While the choice of the halo model only introduces an uncertainty on the 
$J$-factor of the order of 10\% in the case of DM decay, for DM annihilations the $J$-factors vary by more than an 
order of magnitude. In addition, the lack of a precise knowledge of parameters like $R_\odot$, $\alpha$ or $r_s$ 
introduces a sizeable uncertainty on the $J$-factor even for individual halo profiles. For DM decay this uncertainty 
is similar in size to the uncertainty introduced by the choice of the halo model. For DM annihilation this 
uncertainty is also significant, but less important than the choice between a cuspy or a cored halo profile. 
In addition, the uncertainty on the local DM density enters linearly (decay) or quadratically (annihilation) 
in the calculation of the $J$-factors as detailed in table~\ref{tab:jfactors}.
For definiteness we adopted the Einasto profile with $\rho_{\odot}\simeq0.4\,\text{GeV}\,\text{cm}^{-3}$ as our
baseline model when optimising our ROIs and presenting the main results of this
paper, but rescaling the results for other profiles and other values of $\rho_{\odot}$ using the values given in
table~\ref{tab:jfactors} is straightforward.\footnote{Note that while the optimal ROI for the case of DM decay is practically independent of the DM density profile, for the case of DM annihilation the optimal ROI has a fairly strong dependence on the choice of the halo model. Therefore, although the entries in table~\ref{tab:jfactors} allow rescaling the limits to other halo models, these limits are not necessarily optimal since the ROIs adopted in our analysis are optimised for an Einasto profile.}
%
% calculated these numbers with MathematicaPlots.nb
\begin{table}[t]
\centering
\begin{tabular}{ccccccccc}
\hline
\hline
&& \multicolumn{3}{c}{Decay} && \multicolumn{3}{c}{Annihilation} \\
Profile && ROI & $\Delta\Omega_\text{ROI}$ & $J$-factor && ROI & $\Delta\Omega_\text{ROI}$ & $J$-factor \\
        &&  & sr & \hspace*{-0.5em}($10^{22}\,$GeV\,cm$^{-2}$)\hspace*{-0.5em} &&  & sr & \hspace*{-0.5em}($10^{22}\,$GeV$^2$\,cm$^{-5}$)\hspace*{-0.5em} \\
        &&	&	& \hspace*{-0.5em}$\times \left(\frac{\rho_\odot}{0.4\,\text{GeV\,cm}^{-3}}\right)$\hspace*{-0.5em} &&	&	& \hspace*{-0.5em}$\times \left(\frac{\rho_\odot}{0.4\,\text{GeV\,cm}^{-3}}\right)^2$\hspace*{-0.5em} \\
\hline
\textbf{Einasto} && ROI$_{\text{pol}}$ & 1.68 & 2.89 && ROI$_{\text{cen}}$ & 0.121 & 8.89 \\
NFW && ROI$_{\text{pol}}$ & 1.68 & 2.96 && ROI$_{\text{cen}}$ & 0.121 & 4.81 \\
Isothermal && ROI$_{\text{pol}}$ & 1.68 & 3.09 && ROI$_{\text{cen}}$ & 0.121 & 1.32 \\
Burkert && ROI$_{\text{pol}}$ & 1.68 & 2.75 && ROI$_{\text{cen}}$ & 0.121 & 0.50 \\
\hline
\hline
\end{tabular}
\caption{\label{tab:jfactors} Summary of the $J$-factors obtained in the two search regions, for
different DM profiles. Our main results assume the Einasto profile and a local DM density 
$\rho_\odot=0.4\,\text{GeV\,cm}^{-3}$, but a rescaling to other profiles and other values of $\rho_\odot$ is straightforward.}
\end{table}

\section{Other control samples}\label{app:control_samples}

To further test the effects of uncertainties of the effective area and the background modelling we used two 
additional control samples along with the Galactic plane scans described in section~\ref{sec:syst}:
the Earth's limb and the Vela pulsar.  Similarly to the Galactic plane scan, these fits were performed in 0.25$\sigma_E$ energy steps.  Both samples were used extensively in the validation of the \Fermi\/-LAT 
IRFs~\cite{REF:2012.P7Perf}, 
and the Earth's limb has been used as a control sample in previous
line searches~\cite{Bloom,Fermi-LAT:2013uma,Finkbeiner:2012ez}.   
These two samples complement each other well; the pulsed emission from Vela cuts off above a few GeV, while the effects of the Earth's geomagnetic cut-off 
on cosmic rays significantly complicates modelling the \gammaRayHyph\ emission from the Earth's limb below a few GeV. 

\begin{table}[t]
  \centering
  \begin{tabular}{lccc}
    \hline\hline
  Selection & Primary data\  & \ Vela pulsar data \ & Limb data\\
  \hline
  Observation Period End & 2013 Oct.\ 15 & 2013 Aug.\ 8 & 2013 Jan.\ 11 \\
  \ \ Mission Elapsed Time (s)\tablefootnote{\Fermi\ Mission Elapsed Time is defined as seconds since 2001 January 1, 00:00:00 UTC.} & 403509400 & 379556800 & 397631400\\
  Fit Energy range (GeV) & \multicolumn{2}{c}{0.1$-$10} & 3.5$-$10 \\
  ROI & see section~\ref{subsec:ROI} & \multicolumn{1}{|c}{$22^{\circ}$ around Vela} & - \\
  Zenith range (deg) & \multicolumn{2}{c}{$\theta_{\rm z} < 100$\tablefootnote{For the Vela pulsar sample we require that the entire $22^{\circ}$ radius ROI pass 
    the zenith angle selection using the {\emph{gtmktime ScienceTool}}.  This effectively requires $\theta_{\rm z} < 78^{\circ}$ at the centre of the ROI.}} & $111< \theta_{\rm z} < 113$ \\
  Rocking angle range (deg)\tablefootnote{Applied by selecting on 
    \texorpdfstring{\texttt{ROCK\_ANGLE}}{ROCK_ANGLE} 
    with the {\emph{gtmktime ScienceTool}}.} & \multicolumn{2}{c}{$|\theta_{\rm r}| < 52$} & $|\theta_{\rm r}| > 52$  \\
  Data quality cut\tablefootnote{Standard data quality selection:
    \texorpdfstring{\texttt{DATA\_QUAL == 1 \&\& LAT\_CONFIG ==
        1}}{DATA_QUAL == 1 \&\& LAT_CONFIG == 1}  
    with the {\emph{gtmktime ScienceTool}}.} & \multicolumn{2}{c}{yes} & yes \\
  \hline\hline
  \end{tabular}
  \caption{\label{tab:event_samples}Summary table of data selections.
  The observation period for all of the samples began 2008 August 4.}
\end{table}

Table~\ref{tab:event_samples} shows the data selections for the primary data and the 
control samples.  With the exception of the differences listed in table~\ref{tab:event_samples}, 
the initial data preparation for the control samples were identical to
those used from the primary data.  For the Vela pulsar 
control sample we then used the {\emph TEMPO2} package\footnote{\url{http://www.atnf.csiro.au/research/pulsar/tempo2/}}~\citep{REF:2006:TEMPO2}
and a pulsar timing model\footnote{\url{http://fermi.gsfc.nasa.gov/ssc/data/access/lat/ephems/}} 
derived from data taken with the Parkes radio telescope~\cite{REF:2010.VelaII,REF:2010:VelaParkes} 
to assign a phase with respect to the 89 ms pulse period to each \gammaRay.  

For the Vela pulsar control sample, we fit for a line using only the on-pulse data (\gammaRays\ with phases in the ranges $[0.1,0.3]\cup[0.5,0.6]$), 
and used the off-pulse data (\gammaRays\ with phases in the range $[0.7,1.0]$) as a spectral model for the astrophysical emission in the ROI
not associated with the Vela pulsar.  We then modelled on-pulse flux of the Vela pulsar emission using a power law with a hyper-exponential cut-off:
\begin{linenomath}
\begin{equation}\label{eq:VelaModel}
   \mu_{\text{Vela},\,i}=\int_{E_i^-}^{E_i^+} dE\, \mathcal{E}(E)\,  E^{\Gamma} \exp[-(E/E_{\rm c})^{b}]\,,
\end{equation}
\end{linenomath}
and fixed $b = 1$ and $E_{\rm c} = 3$\,GeV, which are slightly different than the values reported in ref.~\cite{REF:2010.VelaII} for the phase-averaged spectrum.

For the Earth's limb control sample, we used a phenomenological model to describe the effect of the Earth's geomagnetic cut-off around 1\,GeV on the 
\gammaRayHyph\ spectrum, 
\begin{linenomath}
\begin{equation}\label{eq:LimbModel}
  \mu_{\text{limb},\,i} =  \int_{E_i^-}^{E_i^+} dE\, \mathcal{E}(E)\,  E^{\Gamma_{1}} [ 1 + ( E / E_{\rm b} )^{(\Gamma_{1}-\Gamma_{2})/\beta} ]^{-\beta},  
\end{equation}
\end{linenomath}
and fixed $\Gamma_{1} = -1.532$, $E_{\rm b} = 370.3$\,MeV and $\beta = 0.7276$. These parameters control the spectrum below the cutoff and the cutoff itself.  However, since we limit our fits in the Limb to $E_{\gamma} > 2$\,GeV (above the cutoff), we fix these parameters.

An example of a fit to the Vela pulsar control samples including the signal and background components is shown in figure~\ref{fig:control_fit_examples}.

\twopanel{t}{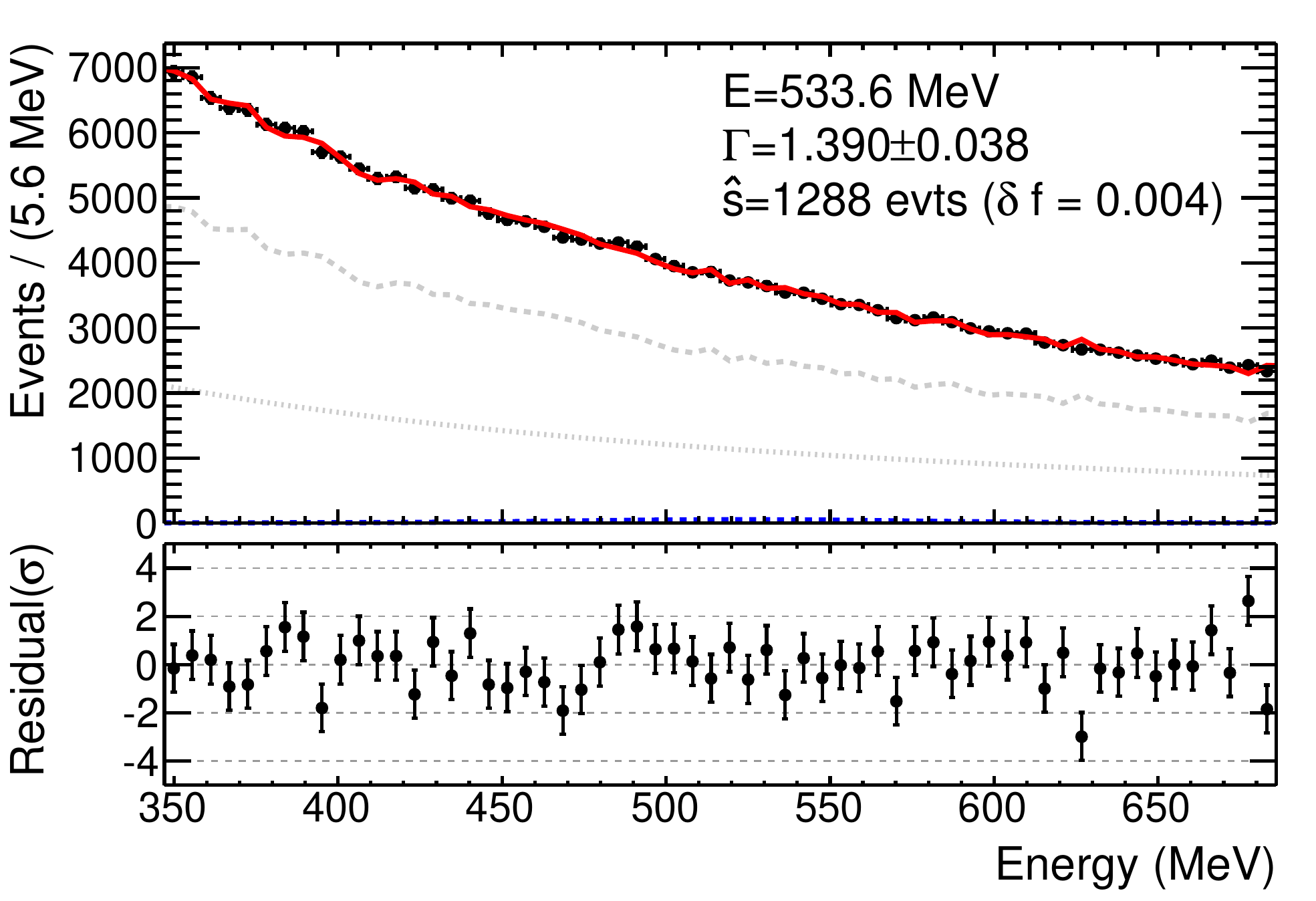}{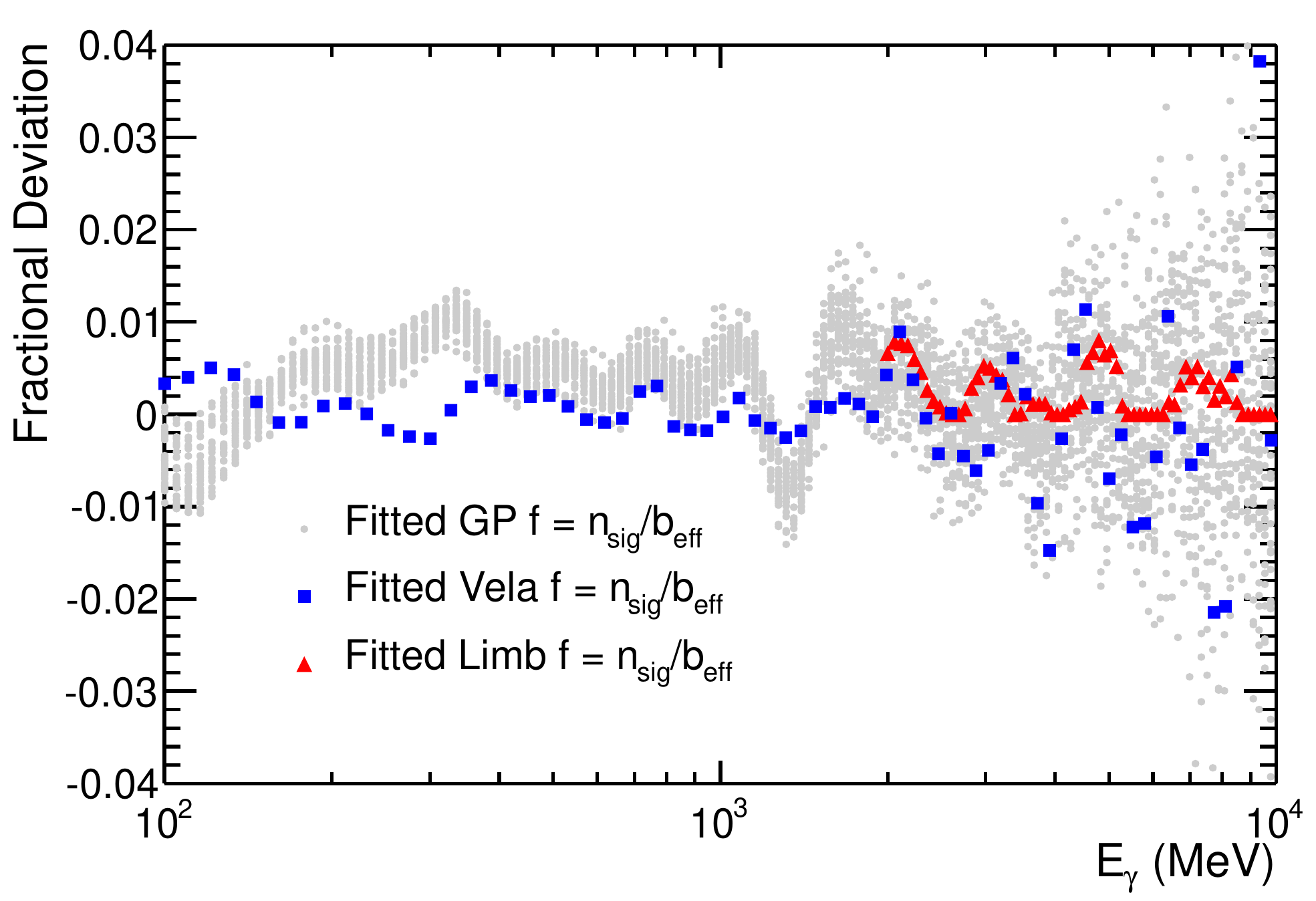}{
  \caption{(left) Fit to the Vela pulsar control sample at $E_{\gamma}=534$\,MeV, showing the models for the off-pulse background (dotted grey line) and the pulsed emission from Vela (dashed grey line); the barely visible dotted blue line shows the signal model.  
    (right) Fractional deviations observed in the Vela pulsar (blue squares), the Earth's limb (red triangles), and the Galactic plane (GP) scan (grey dots).}
  \label{fig:control_fit_examples}
}

The estimates of the fractional residuals from these control samples 
are shown in figure~\ref{fig:control_fit_examples}, and are somewhat smaller than from the scan of the Galactic plane.  Since
the instrumental uncertainties are similar for
all the control samples, this suggests that background modelling
uncertainties are larger for the control regions in the Galactic plane
than for the Vela pulsar and the Earth's limb.  However, notably, from
400\,MeV to 2\,GeV the results from the Vela pulsar appear to track
the results from the Galactic plane control samples, suggesting that
these deviations might be caused by a common systematic effect such as energy-dependent
variations in the effective area.

\section{ROI optimisation}\label{app:ROIs}
As discussed above, the present study is limited by systematic rather than
statistical uncertainties.  We will hence adopt ROIs that optimise
the signal-to-background ratio, rather than the signal-to-noise
ratios which is relevant for minimising statistical noise. We note however that
we find similar results when considering regions with optimal signal-to-noise
ratio. In our search for the best regions, we kept the ROI shape initially
identical to the shape used in the high energy line search in
ref.~\cite{Fermi-LAT:2013uma}.  The shape was defined as a circle of radius $R_0$
centred on the Galactic Centre, excluding part of the disc up to latitude of
$B_0$ while keeping the Galactic Centre in a longitude range $[-L_0,+L_0]$.
Namely, we defined the ROIs as the set of points satisfying the following conditions$\text{:}\quad(\psi < R_0) \quad \text{and} \quad (|b|>B_0 \quad \text{or}
\quad |\ell|<L_0)$, where $\psi$ denotes the angular distance from the Galactic
Centre.  

We derive the number of expected signal events within the ROI from the
baseline Einasto profile in case of both DM annihilation and decay, assuming
uniform exposure of the sky. Note that the overall normalisation does not
matter for this discussion though the results change very little using the \irf{P7REP\_CLEAN}\ exposure.  
We obtain the number of expected background events 
directly from the distribution of measured $\gamma$-rays above 1\,GeV (using
\irf{P7CLEAN} data, though the results are very similar using the
\irf{P7REP\_CLEAN} data).  All three parameters $R_0$, $B_0$ and $L_0$ are then varied within
their physically allowed ranges to find the ROI with maximal
signal-to-background ratio, $n_\text{sig}/b_\text{eff}$. We neglect the impact
of the LAT PSF when evaluating $n_\text{sig}$.  Note
that in cases where both signal and background fluxes lack a strong spatial
dependence in the relevant regions of the sky, the signal-to-noise ratio, in
general, increases with an increasing ROI size.  This is not the case for the
signal-to-background ratio, where the surface factor cancels out.  As a
consequence, we do not find very tight constraints on the size
of the ROIs.  Parameters that yield a very high signal-to-background ratio are
in case of DM decay $R_0=180^\circ$, $L_0 = 0^\circ$ and $B_0=60^\circ$ (we
denote this region by ROI$_\text{pol}$), and in case of DM annihilation $R_0=8^\circ$,
$L_0 = 2^\circ$ and $B_0 = 8^\circ$.  However, since the signal flux in the
second ROI would be severely affected by the broad PSF of the \Fermi\/-LAT at
very low energies, we selected a simple $20^\circ\times20^\circ$ ROI around the
Galactic Centre instead (ROI$_\text{cen}$).  We find that this region reduces the
signal-to-background ratio by about a factor of two with respect to the optimal region obtained when
PSF effects are neglected.

\section{Tabulated limits}
\label{app:tables}

We provide the flux limits for ROI$_{\text{pol}}$ and ROI$_{\text{cen}}$ as well as the limits on the decay lifetime and the
thermal-averaged annihilation cross-section in tables~\ref{tab:limits} and~\ref{tab:limitsCont}. These correspond to the limits presented in figures~\ref{fig:FluxLim} and~\ref{fig:DMLim}.

\newpage

\begin{table}
\footnotesize
\centering
\begin{tabular}{ccccccccc}
\hline
\hline
  & \multicolumn{4}{c}{DM Decay -- ROI$_{\text{pol}}$} & \multicolumn{4}{c}{DM Annihilation -- ROI$_{\text{cen}}$} \\
  & \multicolumn{2}{c}{stat.\ only} & \multicolumn{2}{c}{stat.\ + syst.} & \multicolumn{2}{c}{stat.\ only} & \multicolumn{2}{c}{stat.\ + syst.} \\
 $E_\gamma$ & $\phi_\text{mono}$ & $\tau_{\gamma\nu}$ & $\phi_\text{mono}$ & $\tau_{\gamma\nu}$ & $\phi_\text{mono}$ & $\left\langle\sigma v\right\rangle_{\gamma\gamma}$ & $\phi_\text{mono}$ & $\left\langle\sigma v\right\rangle_{\gamma\gamma}$ \\
 (GeV) & ($\text{cm}^{-2}\text{s}^{-1}$) & (s) & ($\text{cm}^{-2}\text{s}^{-1}$) & (s) & ($\text{cm}^{-2}\text{s}^{-1}$) & (cm$^3$s$^{-1}$) & ($\text{cm}^{-2}\text{s}^{-1}$) & (cm$^3$s$^{-1}$) \\
 & $\times 10^{-8}$ & $\times 10^{28}$ & $\times10^{-8}$ & $\times 10^{28}$ & $\times10^{-8}$ & $\times10^{-30}$ & $\times 10^{-8}$ & $\times10^{-30}$ \\
\hline
0.100 & 1.04 & 111 & 164 & 0.701 & 6.71 & 0.0949 & 115 & 1.63 \\
0.110 & 0.378 & 276 & 106 & 0.988 & 2.25 & 0.0386 & 99.3 & 1.70 \\
0.121 & 0.321 & 296 & 85.8 & 1.11 & 2.70 & 0.0560 & 91.7 & 1.90 \\
0.133 & 0.369 & 234 & 79.2 & 1.09 & 6.60 & 0.166 & 88.0 & 2.21 \\
0.146 & 0.445 & 177 & 74.0 & 1.06 & 9.00 & 0.272 & 82.3 & 2.49 \\
0.161 & 1.60 & 44.7 & 78.1 & 0.918 & 15.2 & 0.554 & 80.9 & 2.95 \\
0.176 & 16.9 & 3.87 & 83.4 & 0.783 & 19.2 & 0.841 & 78.8 & 3.45 \\
0.193 & 18.7 & 3.20 & 75.0 & 0.795 & 19.6 & 1.03 & 73.6 & 3.88 \\
0.211 & 18.8 & 2.89 & 65.5 & 0.832 & 15.3 & 0.962 & 64.5 & 4.07 \\
0.231 & 10.6 & 4.71 & 50.4 & 0.988 & 13.6 & 1.02 & 58.3 & 4.40 \\
0.252 & 5.61 & 8.13 & 42.0 & 1.08 & 13.1 & 1.18 & 54.8 & 4.94 \\
0.276 & 6.57 & 6.36 & 39.6 & 1.05 & 15.2 & 1.63 & 55.0 & 5.91 \\
0.301 & 12.2 & 3.14 & 41.9 & 0.914 & 16.8 & 2.14 & 54.2 & 6.93 \\
0.327 & 12.7 & 2.77 & 40.0 & 0.878 & 17.4 & 2.63 & 52.8 & 8.00 \\
0.356 & 8.60 & 3.75 & 33.5 & 0.963 & 15.5 & 2.78 & 48.5 & 8.71 \\
0.387 & 2.49 & 12.0 & 25.0 & 1.19 & 10.3 & 2.18 & 40.4 & 8.55 \\
0.420 & 1.17 & 23.4 & 21.4 & 1.28 & 6.85 & 1.71 & 34.4 & 8.58 \\
0.455 & 2.42 & 10.5 & 20.3 & 1.25 & 6.66 & 1.95 & 31.6 & 9.24 \\
0.492 & 2.96 & 7.89 & 18.4 & 1.27 & 5.47 & 1.87 & 27.7 & 9.49 \\
0.532 & 3.26 & 6.63 & 16.4 & 1.32 & 5.23 & 2.09 & 25.1 & 10.0 \\
0.574 & 2.38 & 8.41 & 14.0 & 1.43 & 4.56 & 2.13 & 22.5 & 10.5 \\
0.619 & 1.59 & 11.7 & 11.9 & 1.57 & 4.60 & 2.49 & 20.9 & 11.3 \\
0.666 & 2.16 & 7.99 & 11.2 & 1.54 & 3.46 & 2.17 & 18.2 & 11.4 \\
0.716 & 3.49 & 4.61 & 11.5 & 1.40 & 4.52 & 3.28 & 17.8 & 12.9 \\
0.769 & 2.24 & 6.68 & 9.33 & 1.60 & 4.33 & 3.62 & 16.7 & 14.0 \\
0.824 & 1.32 & 10.6 & 7.63 & 1.83 & 2.33 & 2.24 & 13.8 & 13.3 \\
0.883 & 0.869 & 15.0 & 6.41 & 2.03 & 1.46 & 1.61 & 11.9 & 13.1 \\
0.945 & 0.628 & 19.4 & 5.50 & 2.21 & 2.35 & 2.97 & 11.7 & 14.8 \\
1.01 & 0.637 & 17.9 & 4.98 & 2.29 & 4.17 & 6.03 & 12.7 & 18.4 \\
1.08 & 0.877 & 12.1 & 4.74 & 2.25 & 4.54 & 7.49 & 12.6 & 20.8 \\
1.15 & 0.972 & 10.3 & 4.42 & 2.26 & 3.19 & 6.01 & 9.87 & 18.6 \\
1.23 & 0.201 & 46.6 & 2.99 & 3.13 & 0.319 & 0.683 & 5.99 & 12.8 \\
1.31 & 0.124 & 70.7 & 2.39 & 3.68 & 0.0950 & 0.231 & 3.96 & 9.62 \\
1.40 & 0.216 & 38.2 & 2.66 & 3.10 & 0.121 & 0.334 & 4.13 & 11.4 \\
1.48 & 0.698 & 11.1 & 3.07 & 2.53 & 0.528 & 1.64 & 5.33 & 16.6 \\
1.58 & 1.07 & 6.80 & 3.27 & 2.23 & 2.12 & 7.44 & 6.75 & 23.7 \\
1.67 & 0.761 & 9.04 & 2.61 & 2.63 & 2.17 & 8.61 & 6.09 & 24.1 \\
1.77 & 0.695 & 9.34 & 2.30 & 2.82 & 1.92 & 8.54 & 5.63 & 25.1 \\
\hline
\hline
\end{tabular}
\caption{\label{tab:limits}95\% CL upper limits on \gammaRayHyph\ fluxes, lower limits
    on DM lifetimes and upper limits on DM annihilation cross sections. We use
    the Einasto profile to translate flux into lifetime and cross section
    limits, see appendix~\ref{app:halo_profile}.}
\end{table}

\begin{table}
\footnotesize
\centering
\begin{tabular}{ccccccccc}
\hline
\hline
  & \multicolumn{4}{c}{DM Decay -- ROI$_{\text{pol}}$} & \multicolumn{4}{c}{DM Annihilation -- ROI$_{\text{cen}}$} \\
  & \multicolumn{2}{c}{stat.\ only} & \multicolumn{2}{c}{stat.\ + syst.} & \multicolumn{2}{c}{stat.\ only} & \multicolumn{2}{c}{stat.\ + syst.} \\
 $E_\gamma$ & $\phi_\text{mono}$ & $\tau_{\gamma\nu}$ & $\phi_\text{mono}$ & $\tau_{\gamma\nu}$ & $\phi_\text{mono}$ & $\left\langle\sigma v\right\rangle_{\gamma\gamma}$ & $\phi_\text{mono}$ & $\left\langle\sigma v\right\rangle_{\gamma\gamma}$ \\
 (GeV) & ($\text{cm}^{-2}\text{s}^{-1}$) & (s) & ($\text{cm}^{-2}\text{s}^{-1}$) & (s) & ($\text{cm}^{-2}\text{s}^{-1}$) & (cm$^3$s$^{-1}$) & ($\text{cm}^{-2}\text{s}^{-1}$) & (cm$^3$s$^{-1}$) \\
 & $\times 10^{-8}$ & $\times 10^{28}$ & $\times10^{-8}$ & $\times 10^{28}$ & $\times10^{-8}$ & $\times10^{-30}$ & $\times 10^{-8}$ & $\times10^{-30}$ \\
\hline
1.88 & 0.533 & 11.5 & 2.00 & 3.06 & 1.63 & 8.16 & 4.94 & 24.7 \\
1.99 & 0.464 & 12.5 & 1.83 & 3.16 & 1.19 & 6.66 & 4.12 & 23.1 \\
2.10 & 0.687 & 7.96 & 1.95 & 2.80 & 0.853 & 5.34 & 3.57 & 22.3 \\
2.22 & 0.776 & 6.68 & 1.92 & 2.70 & 1.17 & 8.18 & 3.68 & 25.6 \\
2.34 & 0.673 & 7.30 & 1.77 & 2.78 & 1.39 & 10.8 & 3.71 & 28.8 \\
2.47 & 0.212 & 21.9 & 1.15 & 4.07 & 0.923 & 7.95 & 2.97 & 25.6 \\
2.60 & 0.0913 & 48.5 & 0.753 & 5.88 & 0.136 & 1.30 & 1.72 & 16.5 \\
2.74 & 0.123 & 34.2 & 0.786 & 5.35 & 0.131 & 1.39 & 1.59 & 16.8 \\
2.88 & 0.432 & 9.24 & 1.16 & 3.45 & 0.603 & 7.09 & 2.17 & 25.5 \\
3.03 & 0.712 & 5.33 & 1.37 & 2.77 & 0.953 & 12.4 & 2.32 & 30.3 \\
3.19 & 0.421 & 8.55 & 1.02 & 3.54 & 0.637 & 9.19 & 1.86 & 26.9 \\
3.36 & 0.347 & 9.86 & 0.871 & 3.93 & 0.301 & 4.82 & 1.40 & 22.3 \\
3.54 & 0.195 & 16.7 & 0.646 & 5.04 & 0.216 & 3.82 & 1.20 & 21.3 \\
3.72 & 0.0980 & 31.6 & 0.457 & 6.76 & 0.160 & 3.14 & 1.02 & 20.0 \\
3.91 & 0.0789 & 37.3 & 0.399 & 7.37 & 0.325 & 7.02 & 1.17 & 25.4 \\
4.11 & 0.104 & 26.9 & 0.434 & 6.45 & 0.505 & 12.1 & 1.33 & 31.8 \\
4.32 & 0.216 & 12.3 & 0.541 & 4.92 & 0.467 & 12.3 & 1.20 & 31.6 \\
4.54 & 0.196 & 12.9 & 0.488 & 5.20 & 0.532 & 15.5 & 1.15 & 33.4 \\
4.77 & 0.147 & 16.4 & 0.396 & 6.09 & 0.482 & 15.5 & 1.03 & 33.2 \\
5.01 & 0.196 & 11.7 & 0.440 & 5.22 & 0.408 & 14.5 & 0.918 & 32.6 \\
5.26 & 0.210 & 10.4 & 0.449 & 4.87 & 0.273 & 10.7 & 0.705 & 27.6 \\
5.52 & 0.152 & 13.7 & 0.353 & 5.90 & 0.108 & 4.66 & 0.461 & 19.9 \\
5.80 & 0.123 & 16.1 & 0.314 & 6.32 & 0.0735 & 3.49 & 0.369 & 17.6 \\
6.09 & 0.107 & 17.7 & 0.273 & 6.93 & 0.153 & 8.00 & 0.488 & 25.5 \\
6.39 & 0.167 & 10.8 & 0.310 & 5.80 & 0.281 & 16.2 & 0.599 & 34.6 \\
6.70 & 0.295 & 5.82 & 0.444 & 3.87 & 0.276 & 17.5 & 0.547 & 34.8 \\
7.03 & 0.301 & 5.44 & 0.438 & 3.74 & 0.181 & 12.7 & 0.407 & 28.4 \\
7.37 & 0.0763 & 20.4 & 0.188 & 8.29 & 0.146 & 11.2 & 0.355 & 27.3 \\
7.73 & 0.0657 & 22.6 & 0.158 & 9.40 & 0.174 & 14.7 & 0.367 & 31.0 \\
8.11 & 0.0812 & 17.5 & 0.174 & 8.16 & 0.256 & 23.8 & 0.473 & 43.9 \\
8.50 & 0.153 & 8.82 & 0.226 & 5.99 & 0.270 & 27.6 & 0.421 & 43.0 \\
8.91 & 0.116 & 11.1 & 0.198 & 6.51 & 0.146 & 16.4 & 0.271 & 30.4 \\
9.34 & 0.0868 & 14.2 & 0.170 & 7.24 & 0.0810 & 10.0 & 0.192 & 23.7 \\
9.80 & 0.134 & 8.78 & 0.200 & 5.86 & 0.0535 & 7.27 & 0.158 & 21.4 \\
\hline
\end{tabular}
\caption{\label{tab:limitsCont}95\% CL upper limits on \gammaRayHyph\ fluxes, lower
    limits on DM lifetimes and upper limits on DM annihilation cross sections
    (continued).  We use
    the Einasto profile to translate flux into lifetime and cross section
    limits, see appendix~\ref{app:halo_profile}.}
\end{table}

\clearpage

\bibliographystyle{JHEP}
\bibliography{LowEnergyLines}

\end{document}